\documentclass[aps,prb,reprint,preprintnumbers,amsmath,amssymb,floatfix]{revtex4-1}

\usepackage{longtable}
\usepackage{morefloats}
\usepackage[dvips]{graphicx}
\usepackage{color}
\usepackage{epsfig,graphicx,amsfonts,amsbsy}
\usepackage{amsmath,amsfonts,amsthm,amssymb}
\usepackage{appendix}
\usepackage{makeidx}
\usepackage{url}
\usepackage{verbatim}
\usepackage[bookmarksnumbered,pdfpagelabels=true,plainpages=false,colorlinks=true,linkcolor=blue,citecolor=blue,urlcolor=blue]{hyperref}
\usepackage[rightcaption]{sidecap}
\usepackage{array}
\usepackage{booktabs}
\usepackage{multirow}
\usepackage{tabularx}
\usepackage{cancel,soul,ulem}

\newcommand{\bs}[1]{\boldsymbol{#1}}

\makeatletter

\begin{document}

\title{Spin Transport in Thick Insulating Antiferromagnetic Films}
\author{Roberto E. Troncoso$^{1}$}
\author{Scott A. Bender$^{2}$}
\author{Arne Brataas$^{1}$}
\author{Rembert A. Duine$^{1,2,3}$}
\affiliation{$^{1}$Center for Quantum Spintronics, Department of Physics, Norwegian University of Science and Technology, NO-7491 Trondheim, Norway}
\affiliation{$^{2}$Utrecht University, Leuvenlaan 4, 3584 CE Utrecht, The Netherlands}
\affiliation{$^{3}$Department of Applied Physics, Eindhoven University of Technology,
P.O. Box 513, 5600 MB Eindhoven, The Netherlands}

\begin{abstract}
Spin transport of magnonic excitations in uniaxial insulating antiferromagnets (AFs) is investigated. In linear response to spin biasing and a temperature gradient, the spin transport properties of normal-metal--insulating antiferromagnet--normal-metal heterostructures are calculated. We focus on the thick-film regime, where the AF is thicker than the magnon equilibration length. This regime allows the use of a drift-diffusion approach, which is opposed to the thin-film limit considered by Bender {\it et al.} 2017, where a stochastic approach is justified. We obtain the temperature- and thickness-dependence of the structural spin Seebeck coefficient $\mathcal{S}$ and magnon conductance $\mathcal{G}$. In their evaluation we incorporate effects from field- and temperature-dependent spin conserving inter-magnon scattering processes. Furthermore, the interfacial spin transport is studied by evaluating the contact magnon conductances in a microscopic model that accounts for the sub-lattice symmetry breaking at the interface.  We find that while inter-magnon scattering does slightly suppress the spin Seebeck effect, transport is generally unaffected, with the relevant spin decay length being determined by non-magnon-conserving processes such as Gilbert damping. In addition, we find that while the structural spin conductance may be enhanced near the spin flip transition, it does not diverge due to spin impedance at the normal metal|magnet interfaces.
\end{abstract}

\maketitle
\section{Introduction}

Spin-wave excitations in magnetic materials are a cornerstone in spintronics for the transport of spin-angular momentum \cite{Antiferromagnets2016,Baltz2018rmp}. The usage of antiferromagnetic materials has gained a renewed interest due to their high potential for practical applications. The most attractive properties of antiferromagnets (AFs) are the lack of stray fields and the fast dynamics that can operate in the THz frequency range \cite{Jungwirth2018}. Those attributes have the potential to tackle current technological bottlenecks, like the absence of practical solutions to generate and detect electromagnetic waves in the spectrum ranging from 0.3 THz to 30 THz (the terahertz gap) \cite{Baltz2018rmp}. Nevertheless, the control and access to the high-frequency response of AFs is challenging. New proposals circumvent one of these obstacles by manipulating metallic AFs with charge currents, through the so-called spin-orbit torques \cite{zelezny2014,wadley2016,Mn2Au2018}. Antiferromagnetic \textit{insulators}, however, offer a compelling alternative since the Joule heating caused by moving electrons is absent. In such systems, the study of transport instead focuses on their magnetic excitations.

In insulating AFs the spin-angular momentum is transferred by their quantized low-energy excitations, i.e., magnons. Since the AF in its groundstate is composed of two collinear magnetic sublattices, magnons carry opposite spin angular momentum. The transport of magnons has been experimentally achieved through the longitudinal spin Seebeck effect in AF$|$NM \cite{Wuprl2016,Sekiprl2015,HolandaAPL2017,ShiomiPRB2017,RibeiroPRB2019,FlebusPRB2019,FlebusPRB2019b} (NM, normal metal) and FM$|$AF$|$NM \cite{WangPRL2014,LinPRL2016,PrakashPRB2016,Qiu2018,Sakimura2019} (FM, ferromagnets) heterostructures, in which magnons were driven by a thermal gradient across the AF. Alternatively, thermal injection of magnons in AFs has been studied \cite{BenderPRL2017,Tatara2019} by a spin accumulation at the contact with adjacent metals. In addition, it was shown that thermal magnon transport takes place at zero spin bias when the sublattice symmetry is broken at the interfaces, e.g., induced by interfacial magnetically uncompensated AF order \cite{chengprl2014}. Complementarily, coherent spin transport induced by spin accumulation has been earlier considered and predicted to result in spin superfluidity \cite{TakeiPRB2014,QaiumzadehPRL2017} or Bose-Einstein condensates of magnons \cite{EirikPRB2017}. Recently, it has been shown via non-local spin transport measurements that magnons remarkably propagate at long distances in insulating AFs; $\alpha$-Fe$_2$O$_3$ \cite{Lebrun2018}, Cr$_2$O$_3$ \cite{Yuaneaat1098}, MnPS$_3$ \cite{XingPRX2019} and also in NiO via spin-pumping experiments with YIG\cite{WangPRL2014}. Their exceptional transport properties, as well as those reported in Refs. \cite{Wuprl2016,Sekiprl2015,HolandaAPL2017,ShiomiPRB2017,RibeiroPRB2019}, are governed by the spin conductance and spin Seebeck coefficients. Rezende $et~al.$ \cite{RezendePRB2016b} discussed theoretically the spin Seebeck effect in AFs in contact with a normal metal. They obtain the Seebeck coefficient in terms of temperature and magnetic field, finding a good qualitative agreement with measurements in MnF$_2$/Pt \cite{Wuprl2016}. In addition, it was found that magnon scattering processes affect significantly the spin Seebeck coefficient. Hitherto there has been no complete studies on the underlying mechanism for spin transport coefficients, e.g., their thickness-, temperature- and field-dependence, effects derived from magnon-magnon interactions or when the sublattice symmetry is broken at the interfaces.

\begin{figure}[h!]
	\includegraphics[width=1.15\linewidth,clip=]{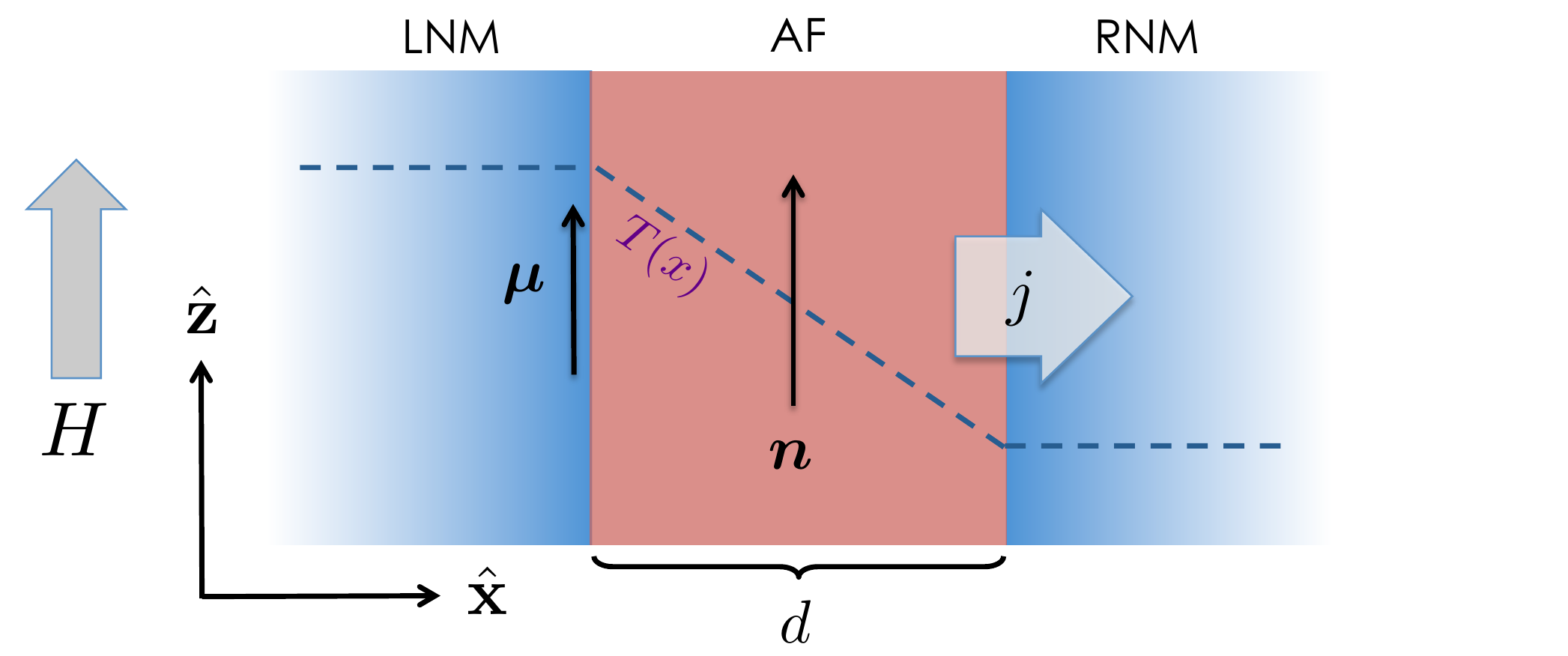}
	\caption{A normal-metal--insulating antiferromagnet--normal-metal heterostructure. An external field ${H}$ is applied along the $z$ direction. A spatially dependent temperature $T(x)$ and a spin bias ${\mathbf \mu}$ is considered. As a result, a magnon spin current $j$ flows through the AF of thickness $d$.}
	\label{sch}
\end{figure}

In this work, we describe spin transport though a left normal-metal--insulating antiferromagnet--right normal-metal (LNM$|$AF$|$RNM) heterostructure. As depicted in Fig. \ref{sch}, magnon transport is driven by either a temperature gradient or spin biasing. We focus on the thick-film limit, where the thickness $d$ of the AF is greater than the internal equilibration length $l$ for the magnon gas. This limit implies a diffusive regime where magnons are in a local equilibrium described by a local temperature and chemical potential. This is in contrast to our earlier, stochastic treatment of thin-films ($d\ll l$) where spin waves do not establish a local equilibrium \cite{BenderPRL2017}. Specifically, we study the spin transport by evaluating, via a phenomenological theory, the structural spin Seebeck coefficient $\mathcal{S}$ and magnon conductance $\mathcal{G}$. Furthermore, we investigate their temperature- and magnetic-field dependence by computing the interfacial conductance coefficients in a microscopic model for the NM$|$AF interface and evaluating the various coefficients using a Boltzmann approach. 

The paper is outlined as follows. In Sec. \ref{sec:model}, we introduce the microscopic Hamiltonian for the bulk AF and its interaction with the metallic contacts.  In Sec. \ref{sec:phenomenology}, we formulate the  phenomenological spin diffusion model, including scattering between magnon branches, and obtain expressions for the structural Seebeck coefficient and magnon conductance. In Sec. \ref{sec:microscopictheory}, we compute the coefficients for interfacial magnon transport from the microscopic model for the contacts. Based on this result, we estimate bulk transport coefficients assuming the interaction parameters are field and temperature independent. We conclude in Sec. \ref{sec:con-disc} with a discussion of our results. In the appendices, we detail various technical aspects of the calculations.

\section{Model}\label{sec:model}
We begin by defining the microscopic model for the LNM$|$AF$|$RNM heterostructure. The total Hamiltonian is $\hat{H}=\hat{H}_{\text{AF}}+\hat{H}_{I}+\hat{H}_{e}$, where $\hat{H}_{AF}$ describes the AF spin system while $\hat{H}_{I}$ represents their interfacial contact with the normal metals. The Hamiltonian $\hat{H}_{e}$ describes the electronic states at the left- and right-lead. The coupling with LNM and RNM is modeled by a simple interfacial exchange Hamiltonian,
\begin{equation}\label{eq:contactHamiltonian}
\hat{H}_{I}=-\int d{\bf x}\sum_{\mathbf{i}}J_{\mathbf{i}}\rho_{\bf i}\left(\bf x\right)\hat{\mathbf{s}}_{\mathbf{i}}\cdot\hat{\mathbf{S}}\left(\mathbf{x}\right),
\end{equation}
where $J_{\bf i}$ is the exchange coupling between the electronic spin density $\hat{\bf S}({\bf x})$ and the localized spin operator $\hat{\bf s}_{\bf i}$ at site ${\bf i}$ that labels the lattice along the interface.  Here $\rho_{\bf i}\left(\bf x\right)$ is the density of the localized AF electron orbital representing effective spin densities at the interface. We will return to the study of $\hat{H}_I$ in Sec. \ref{sec:microscopictheory} to determine the contact spin conductance.

The AF spin Hamiltonian is introduced by labelling each square sublattice site by the position $\mathbf{i}$. The nearest-neighbour Hamiltonian is
\begin{equation}\label{spinHamiltonian}
\hat{H}_{AF}=J\sum_{\left\langle \mathbf{i}\mathbf{j}\right\rangle }\hat{\mathbf{s}}_{\mathbf{i}}\cdot\hat{\mathbf{s}}_{\mathbf{j}}-H\sum_{\mathbf{i}}\hat{s}_{\mathbf{i}z}+\frac{\kappa}{2s}\sum_{\mathbf{i}}\left(\hat{\mathbf{s}}_{\mathbf{i}x}^{2}+\hat{\mathbf{s}}_{\mathbf{i}y}^{2}\right),
\end{equation}
with $\hat{\mathbf{s}}_\mathbf{i}$ the spin operator at site $\mathbf{i}$, $J>0$ the antiferromagnetic exchange biasing, $H$ the magnetic field, and $\kappa$ the uniaxial easy-axis anisotropy. We are interested in small spin fluctuations (magnons) around the collinear bipartite ground state. The latter is the relevant ground state to expand around  for magnetic fieds below the spin-flop field $H_{\text{sf}}$. Magnons are introduced	 by the Holstein-Primakoff transformation \cite{HolsteinPR1940},	
\begin{subequations}
	\begin{align}
	\hat{\mathbf{s}}_{\mathbf{i}z}&=s-\hat{a}_{\mathbf{i}}^{\dagger}\hat{a}_{\mathbf{i}},\,\,\,\,\,\,\hat{\mathbf{s}}_{\mathbf{i}-}=\hat{a}_{\mathbf{i}}^{\dagger}\sqrt{2s-\hat{a}_{\mathbf{i}}^{\dagger}\hat{a}_{\mathbf{i}}},
	\label{hpa}
	\\
	\hat{\mathbf{s}}_{\mathbf{i}z}&=-s+\hat{b}_{\mathbf{i}}^{\dagger}\hat{b}_{\mathbf{i}},\,\,\,\,\,\,\hat{\mathbf{s}}_{\mathbf{i}-}=\sqrt{2s-\hat{b}_{\mathbf{i}}^{\dagger}\hat{b}_{\mathbf{i}}}\hat{b}_{\mathbf{i}},
	\label{hpb}
	\end{align}
\end{subequations}
and $\hat{\mathbf{s}}_{\mathbf{i}+}=\hat{\mathbf{s}}^{\dagger}_{\mathbf{i}-}$, when $\mathbf{i}$ belongs to  sublattice $a$ and $b$, respectively. We expand the spin Hamiltonian, Eq. (\ref{spinHamiltonian}), in powers of magnon operators that includes magnon-magnon interactions, up to the fourth order, $\hat{H}_{AF}=\hat{H}^{(2)}_{AF}+\hat{H}^{(4)}_{AF}$.  To lowest order in $s$, excitations of $\hat{H}^{(2)}_{AF}$ are diagonalized through the Bogoliubov transformation (see Appendix \ref{app:scattamp} for definition), by the operators $\hat{\alpha}_\mathbf{q}$ and $\hat{\beta}_\mathbf{q}$ that carry spin angular momentum $+\hbar \hat{\mathbf{z}}$ and $-\hbar \hat{\mathbf{z}}$ respectively,
\begin{equation}\label{eq:diagHam}
\hat{H}^{(2)}_{AF}=\sum_{\mathbf{q}}\left[\epsilon_\alpha(\mathbf{q})\hat{\alpha}_{\mathbf{q}}^{\dagger}\hat{\alpha}_{\mathbf{q}}+\epsilon_\beta(\mathbf{q})\hat{\beta}_{\mathbf{q}}^{\dagger}\hat{\beta}_{\mathbf{q}}\right].
\end{equation}
We refer to the magnons described by the operator $\alpha$ ($\beta$) as $\alpha$-($\beta$)magnons, respectively. The dispersion relation is $\epsilon_{\alpha,\beta}(\mathbf{q})=\pm H+\sqrt{\left(6Js\right)^{2}\left(1-\gamma_{\mathbf{q}}^{2}\right)+H_{\text{c}}^2}$
in a 3-dimensional lattice, where $\pm$ stands for the $\alpha$- and $\beta$-magnon branch, respectively. Here, $H_{\text{c}}^2 \equiv \kappa^{2}+2\kappa 6Js$ is the critical field corresponding to the spin-flop transition, while $\gamma_\mathbf{q}=(1/3)\sum_{n=1}^3\mathrm{cos}(q_n\mathfrak{a})$, where $\mathfrak{a}$ is the lattice spacing. Magnon-magnon interactions are represented by the interacting Hamiltonian $\hat{H}^{(4)}_{AF}$. In the diagonal basis, the interacting Hamiltonian becomes a lengthy expression that is detailed in Eq. (\ref{eq:magmaghamiltonian}) (Appendix \ref{app:scattamp}). It consists of nine different scattering processes among $\alpha$- and $\beta$-magnons. Some of these processes allow for the exchange of population of $\alpha$- and $\beta$-magnons, see Fig. \ref{feynmagnons}. 

\section{Spin transport: phenomenological theory}\label{sec:phenomenology}
We now outline the phenomenological spin transport theory for magnons across the LNM$|$AF$|$RNM heterostructure. In the subsections that follow, we estimate the structural spin Seebeck coefficient and structural magnon conductances. The basic assumption is that the equilibration length for magnon-magnon interactions is much shorter than the system length $d$, so that the two magnon gases are parametrized by local chemical potentials $\mu_\alpha$ and $\mu_\beta$ and temperatures $T_\alpha$ and $T_\beta$. In keeping with our treatment of ferromagnets \cite{BenderPRB2015}, we assume that strong, inelastic spin-preserving processes fix the local magnon temperatures to that of the local phonon temperature. This assumption is reasonable since the rate at which magnon temperature equilibrates with the phonon bath is dominated by both magnon-conserving and magnon-nonconserving scattering processes\cite{CornelissenPRB2016}. Thus, the magnon temperature reaches its equilibrium faster than the magnon chemical potential. The local phonon temperature, in turn, is assumed to be linear across the AF, and to be equal to the electronic temperatures in each of the metallic leads. Only the magnon chemical potentials $\mu_\alpha$ and $\mu_\beta$ are then left to be determined.

We then express phenomenologically the spin conservation laws in terms of the chemical potentials. Its microscopic derivation can be established from the Boltzmann equation as is explained in Appendix \ref{app:scattlength}. Defining the magnon densities $n_\alpha$ and $n_\beta$, these read
\begin{subequations}
\begin{align}
\label{bulkeom1}
\dot{n}_{\alpha}+\nabla\cdot\mathbf{j}_{\alpha}&=-r_{\alpha}\mu_{\alpha}-g_{\alpha\alpha}\mu_{\alpha}-g_{\alpha\beta}\mu_{\beta},\\
\dot{n}_{\beta}+\nabla\cdot\mathbf{j}_{\beta}&=-r_{\beta}\mu_{\beta}-g_{\beta\alpha}\mu_{\alpha}-g_{\beta\beta}\mu_{\beta}\, .
\label{bulkeom2}
\end{align}
\end{subequations}
Here, $r_i$ describes relaxation of spin into the lattice resulting from inelastic magnon-phonon interactions that do not conserve magnon number. In addition, $g_{ij}$ describes inelastic spin-conserving processes that accounts for, e.g., magnon-magnon and magnon-phonon scattering, where the total number of magnons $n_\alpha+n_\beta$ may change but the spin $\sim n_\alpha-n_\beta$ is constant. In what follows, the coefficients $g_{ij}$, by assumption, have their origin in the coupling between magnons. The currents of $\alpha$- and $\beta$-magnons, denoted as $\mathbf{j}_\alpha$ and $\mathbf{j}_\beta$, are given by $\mathbf{j}_\alpha=-\sigma_\alpha \nabla \mu_\alpha-\varsigma_\alpha \nabla T$ and $\mathbf{j}_\beta=-\sigma_\beta \nabla \mu_\beta-\varsigma_\beta \nabla T$, where $\sigma_{\alpha,\beta}$ and $\varsigma_{\alpha,\beta}$ are the bulk magnon spin conductivities and Seebeck coefficients, respectively. In writing the particle currents in the form above, we have neglected magnon-magnon drag, which stems from magnon-magnon interactions that transfer linear momentum from one magnon band to another in such a way that the total spin current is conserved. Such drag gives rise to cross-terms like $j_\alpha \propto \nabla \mu_\beta$.  We shall simply limit the discussion to the regime in which such momentum scattering in subdominant to e.g. elastic disorder scattering. The bulk continuity equations, Eqs. (\ref{bulkeom1}) and (\ref{bulkeom2}), are complemented by the boundary conditions at the NM$|$AF interfaces on the spin currents $\mathbf{j}^{(s)}_\alpha=\hbar \mathbf{j}_\alpha$ and $\mathbf{j}^{(s)}_\beta=-\hbar \mathbf{j}_\beta$,
\begin{subequations}
\begin{align}
\label{bcL}
\mathbf{x}\cdot \mathbf{j}_{\alpha}^{(s)}(x=-d/2)&=G_\alpha\left[\mu_L-\mu_\alpha(-d/2)\right],\\
\mathbf{x}\cdot \mathbf{j}_{\beta}^{(s)}(x=-d/2)&=G_\beta\left[\mu_L+\mu_\beta(-d/2)\right],\\
\mathbf{x}\cdot \mathbf{j}_{\alpha}^{(s)}(x=d/2)&=-G_\alpha\left[\mu_R-\mu_\alpha(d/2)\right],\\
\mathbf{x}\cdot \mathbf{j}_{\beta}^{(s)}(x=d/2)&=-G_\beta\left[\mu_R+\mu_\beta(d/2)\right],
\label{bcR}
\end{align}
\end{subequations}
with ${\bf x}$ the unit vector along $x$-axis and where we have chosen the left and right interfaces to correspond to the planes $x=-d/2$ and $x=d/2$. Inside the left and right normal metals the respective spin accumulations, the difference between spin-up and spin-down chemical potential, are $\mu_L$ and $\mu_R$. Here $G_{\alpha,\beta}$ are the contact magnon spin conductances of each interface. The contact Seebeck coefficient does not appear, as we are assuming a continuous temperature profile across the structure, i.e., there is no temperature difference between magnons at the interface and normal metal leads. For fixed spin accumulations $\mu_{L/R}$, Eqs. (\ref{bulkeom1}-\ref{bcR}) form a closed set of equations with the parameters $g$, $r_{\alpha,\beta}$, $\varsigma_{\alpha,\beta}$, $\sigma_{\alpha,\beta}$ and $G_{\alpha,\beta}$ to be estimated from microscopic calculations (see Sec. \ref{sec:microscopictheory}).

The inelastic spin-conserving terms $g_{ij}$ can be significantly simplified by additional considerations. Imposing spin conservation one finds that $g_{\alpha\alpha}=g_{\beta \alpha}$ and $g_{\alpha \beta}=g_{\beta \beta}$. This result is obtained from Eqs. (\ref{bulkeom1}-\ref{bcR}) by equating $\dot{n}_{\alpha}-\dot{n}_{\beta}=0$ in the absence of magnon currents and disregarding the relaxation term $r_i$. In addition, we can estimate  the field- and temperature-dependence of the coefficients $g_{\alpha\alpha}$ and $g_{\beta \beta}$, in particular near the spin-flop transition. For this purpose, we use Fermi's golden rule to calculate the transition rate of $\alpha$-magnons ($\beta$-magnons), defined as $\Gamma_{\alpha\beta}$ ($\Gamma_{\beta\alpha}$), that represents the instantaneous leakage of magnons due to the conversion between $\alpha$- and $\beta$-magnons. Among the different scattering processes displayed in Fig. \ref{feynmagnons}, few of them conserve the number of $\alpha$- or $\beta$-magnons and thus do not contribute to the transition rate. As detailed in Appendix \ref{app:scattlength}, we sum over all the scattering rates and find that $\Gamma_{\alpha\beta}=\Gamma_{\beta\alpha}$, which derives as a consequence of conservation of spin-angular momentum. Moreover, and more importantly, up to linear order in the chemical potential $\Gamma_{\alpha\beta}=-g\left(\mu_{\alpha}+\mu_{\beta}\right)$. Therefore, $g_{\alpha\alpha}=g_{\alpha\beta}\equiv g$, meaning that a single scattering rate describes the inelastic spin-conserving process. The coefficient $g$ is expressed in terms of a complex integral, given in Eq. (\ref{eq:gfactor}), that can be estimated in certain limits. In the high temperature regime, where the thermal energy is much higher than the magnon gap, we obtain $g=\left(2\pi N\Omega/{\hbar s^2}\right)\left({k_BT}/{Jsz}\right)^3$ with $\Omega$ a dimensionless integral defined in Appendix \ref{app:scattlength}.

In the steady state limit the magnon chemical potentials are described by Eqs. (\ref{bulkeom1}) and (\ref{bulkeom2}), and are of the general form
$\mu_{\alpha}\sim\mu_{\beta}\sim e^{\pm x/\lambda}$. The collective spin decay length $\lambda$ admits two solutions,
\begin{align}
2 \lambda_1^{-2}\label{eq:collength1}=\Lambda_\alpha^{-2}+\Lambda_\beta^{-2}-\sqrt{4 \lambda_{\beta}^{-2}\lambda_{\alpha}^{-2}+(\Lambda_\alpha^{-2}-\Lambda_\beta^{-2})^2}\\
2 \lambda_2^{-2}\label{eq:collength2}=\Lambda_\alpha^{-2}+\Lambda_\beta^{-2}+\sqrt{4 \lambda_{\alpha}^{-2}\lambda_{\beta}^{-2}+(\Lambda_\alpha^{-2}-\Lambda_\beta^{-2})^2}
\end{align}
where $\lambda_{\beta}^{-2}=g/\sigma_\beta$, $\lambda_{\alpha}^{-2}=g/\sigma_\alpha$,  $\Lambda_\alpha^{-2}=(g+r_\alpha)/\sigma_\alpha$ and  $\Lambda_\beta^{-2}=(g+r_\beta)/\sigma_\beta$. In the absence of magnetic field,
the magnon-bands are degenerate and therefore $\alpha$ and $\beta$ have equal properties. Thus the collective spin diffusion lengths become $\lambda_1^{-2}=r/\sigma$ and $\lambda_2^{-2}=(2g+r)/\sigma$ that differ due to the inelastic spin-conserving scattering ($\sim g$). In the following sections we evaluate the structural spin Seebeck coefficient and structural magnon conductance. We will consider separately two scenarios, a temperature gradient and spin bias across the LNM$|$AF$|$RNM heterostructure in Sec. \ref{sse} and \ref{sec:spinbiasing}, respectively.

\subsection{Spin Seebeck Effect}\label{sse}
Let us assume a linear temperature gradient, with no spin accumulation in the normal metals. We solve for the spin current at the right interface, $j_s=\mathbf{x}\cdot \mathbf{j}_{\alpha}^{(s)}(d/2)+\mathbf{x}\cdot \mathbf{j}_{\beta}^{(s)}(d/2)$ in presence of the temperature gradient $\Delta T$. Then, the spin current flowing through the right interface is related to the thermal gradient by $j_s=\mathcal{S} \Delta T$,
where $\mathcal{S}$ is the structural spin Seebeck coefficient. The general solution for ${\cal S}$ is found in Appendix \ref{app:seebeckcoef} (Eq. \ref{eq:fullstrucSeebeck}). In what follows we examine several regimes of interest.

First, we consider the zero applied magnetic field case, but allow for sublattice symmetry breaking at the normal metal|magnet interfaces. Here, we have 
that the dispersion relations for the $\alpha$- and $\beta$-magnons are identical. Furthermore, the bulk transport properties becomes independent of the magnon band, i.e., $\sigma=\sigma_\alpha=\sigma_\beta$ and $\varsigma=\varsigma_\alpha=\varsigma_\beta$. In this limit one finds
\begin{equation}\label{SbulkSymm}
\mathcal{S}=\frac{2\sigma\varsigma\left(G_{\beta}-G_{\alpha}\right)}{\left(\mathcal{G}_{\alpha1}\mathcal{G}_{\beta2}+\mathcal{G}_{\alpha2}\mathcal{G}_{\beta1}\right)d\lambda_1}\mathrm{Coth}\left[\frac{d}{2\lambda_{1}}\right],
\end{equation}
with $\mathcal{G}_{in}$ the effective conductances defined by $\mathcal{G}_{in}\equiv G_{i}+\left({\sigma_{i}}/{\lambda_{n}}\right)\mathrm{Coth}\left[{d}/{2\lambda_{n}}\right]$ for $i=\alpha,\beta$ and $n=1,2$. We see that ${\cal S}$ is proportional to the bulk spin Seebeck conductivity $\varsigma$. In the absence of symmetry breaking at the interfaces, $G_\alpha= G_\beta$, the spin Seebeck effect vanishes as expected. When there is no magnetic field, it is thus essential to have systems with uncompensated interfaces to get a finite Seebeck effect.

In order to understand the dependence of Eq. (\ref{SbulkSymm}) on the film thickness $d$, it is useful to distinguish two thickness regimes. Let us first introduce a ``thin" film regime, $d\ll d_{in} \equiv\lambda_n \mathrm{Coth}^{-1}(G_i \lambda_n/ \sigma)$ for $n=1,2$ and $i=\alpha,\beta$. In this limit
$\mathcal{G}_{ni}\approx \left({\sigma}/{\lambda_{n}}\right)\mathrm{Coth}\left[d/2\lambda_{n}\right]$. The spin Seebeck coefficient becomes,
\begin{equation}
\mathcal{S}\approx\lambda_{2}\frac{\left(G_{\beta}-G_{\alpha}\right)}{d \mathrm{Coth}\left[d/2\lambda_2\right]}\frac{\varsigma}{\sigma}\, ,
\end{equation}
which in the extreme thin film limit ($d\ll \lambda_2$), becomes independent of $d$, $S\rightarrow (G_\beta-G_\alpha)2\varsigma/\sigma$. This can be understood as the sum of two independent parallel channels, each with effective conductances renormalized by the bulk transport parameters. When $G_i >\sigma/\lambda_n$, we may also define a ``thick" regime ($d\gg d_{in}\equiv \lambda_n \mathrm{Coth}^{-1}(G_i \lambda_n/ \sigma)$ for all $i,n$) in which the contact resistance dominates, i.e., $\mathcal{G}_{in}\approx G_i \,\,\,\,{\rm (thick~film)}$. In this case, one obtains,
\begin{align}
\mathcal{S}\approx\mathrm{Coth}\left[\frac{d}{2\lambda_1}\right]\frac{\sigma\varsigma}{d \lambda_{1}}\left(G_{\alpha}^{-1}-G_{\beta}^{-1}\right),
\end{align}
and $\mathcal{S}\sim \left(\sigma\varsigma/d\lambda_{1}\right)G^{-1}_T$ at long distances $d\gg \lambda_1$. In this case, the net interfacial conductance behaves as the sum of a series spin-channels, each with conductance $G_{\alpha}$ and $G_{\beta}$. Note that as the Seebeck coefficient is defined through the relation $j_s = \mathcal{S} \Delta T$, the $\sim 1/d$-dependence means that $j_s \propto \partial_x T$ is independent of $d$; a Seebeck effect can thus originate for a thick AF due to a difference between the impedances of the magnon-bands just at the interface where the signal is measured.

Second, we consider effects of a finite applied magnetic field. In addition, we assume no symmetry-breaking at interfaces, $G_\alpha=G_\beta=G$.  In the ``thick" film regime, we obtains ${\cal S}\approx -(\varsigma_\alpha-\varsigma_\beta)/d$, which is simply the bulk value of the Seebeck coefficient. However, allowing symmetry breaking at the interfaces we can obtain ${\cal S}$ in the weak coupling regime, i.e., $g\ll r_{\alpha},r_\beta$, corresponding to slow scattering between the magnon branches (compared to Gilbert damping). Expanding the collective spin decay length, Eqs. (\ref{eq:collength1}) and (\ref{eq:collength2}), to linear order in $g/r_{\alpha,\beta}$, we get $\lambda_1\approx \sqrt{\sigma_\beta/r_\beta}\left(1- g/r_\beta\right)$ and $\lambda_2\approx \sqrt{\sigma_\alpha/r_\alpha}\left(1- g/r_\alpha\right)$. This expansion lead to corrections in the structural Seebeck coefficient, $\mathcal{S}\approx \mathcal{S}^{(0)}+\mathcal{O}(g/r)$, where
\begin{equation}\label{eq:S0}
\mathcal{S}^{(0)}=\mathcal{S}^{(0)}_\beta+\mathcal{S}^{(0)}_\alpha=\frac{G_\beta \varsigma_\beta}{d \mathcal{G}^{(0)}_{\beta 1}}-\frac{G_\alpha \varsigma_\alpha}{d \mathcal{G}^{(0)}_{\alpha 2}},
\end{equation}
with $\mathcal{G}_{ni}^{(0)}$ the lowest order of the effective conductances. It is interesting to note that Eq. (\ref{eq:S0}) consists of two completely decoupled parallel channels. In the particular thick film regime ($d\gg d_{in}$), it reduces to $S^{(0)}\approx -(\varsigma_\alpha-\varsigma_\beta)/d$, which is consistent with the result obtained at finite field in the ``thick" film regime and $G_{\alpha}=G_{\beta}$.	Although we allow for symmetry-breaking at the interface here, all of the interfacial properties are washed out in the thick film regime. 

Last, consider the regime in which interactions are strong: $g \gg r_\alpha, r_\beta$ and $d\gg \lambda_2$. Naively, one might expect interactions to greatly reduce the spin Seebeck effect. In fact, one finds that all dependence on $g$ drops out:
\begin{equation}
\mathcal{S}=\frac{G_\alpha+G_\beta}{d}\left[\frac{(\frac{\sigma_\alpha}{\sigma_\beta})^2\varsigma_\beta-\varsigma_\alpha}{(\frac{\sigma_\alpha}{\sigma_\beta})^2\varsigma_\beta+\varsigma_\alpha}\right] \,,
\end{equation}

Thus, even with strong interactions between magnon bands, the spin Seebeck effect becomes independent of $g$ and nonzero. The effects of interband interactions are shown in Fig.~(\ref{fig:structuralSeebeck}a); while there is a slight suppression of the signal, the spin Seebeck effect is qualitatively unchanged by large scattering.

\subsection{Spin biasing}\label{sec:spinbiasing}
Aside from a temperature gradient, a spin current may be generated by means of an electrically driven spin biasing across the spin (usually via the spin Hall effect in a normal metal contact)\cite{Lebrun2018}. To model this, we consider the temperature constant throughout the structure, but a spin accumulation $\boldsymbol{\mu}=\mu \hat{\mathbf{z}}$ is applied at the left interface, giving rise to a spin current $j=\mathcal{G}\mu$ flowing out of the opposite interface, parametrized by the structural conductance coefficient ${\cal G}$. The full steady-state solution to Eqs.~(\ref{bulkeom1}) and~(\ref{bulkeom2}) is given by Eq.~(\ref{full_g}) in Appendix \ref{app:seebeckcoef}. In order to find simple relations for the spin conductance, we focus on three regimes.

First, we consider the case of sublattice symmetry and zero magnetic field
At the interfaces, this entails $G_\alpha=G_\beta=G$. In the bulk, this implies that bulk magnon spin conductivities and Seebeck coefficients are identical for each magnon branch. Here we find that only one collective spin decay length, $\lambda_r=\sqrt{\sigma/r}$, plays a role in transport. One obtains,
\begin{equation}
\label{part_hole_g}
\mathcal{G}=\frac{2G^{2}\sigma/\lambda_{r}}{\left[\sigma^{2}/\lambda_{r}^2+G^{2}\right]\mathrm{sinh}\left(\frac{d}{\lambda_r}\right)+2(\sigma/\lambda_{r})G\mathrm{cosh}\left(\frac{d}{\lambda_r}\right)}.
\end{equation}
(Note that as the field  - or symmetry-breaking at the interfaces - is turned on, the magnon-magnon interactions start to play a role).  In the thin film regime ($d\ll \lambda_r$),  $\mathcal{G}\approx G$, which is just the series conductance of two parallel channels, each with interfacial conductance $G/2$ (due to the two interfaces through which the spin current must pass). In the opposite limit, $d \gg \lambda_r$, we find
\begin{equation}
\mathcal{G}\approx \frac{4(\sigma/\lambda_{r})G^2}{(\sigma/\lambda_{r}+G)^2}e^{-d/{\lambda_r}},
\end{equation}
exhibiting an exponential decay over the distance $\lambda_r$. 

Second, we consider the strongly interacting case where $g \gg r_\alpha, r_\beta$ and $d\gg \lambda_2$. Here,  one finds that while the conductance generally depends on $g$, in this regime the conductance is finite and independent of $g$:
\begin{equation}
\label{large_g_g}
\mathcal{G} = \mathcal{G}_{S}+\mathcal{G}_{B}
\end{equation}
where 
\begin{equation}
\mathcal{G}_{S}=\frac{\left(G_{\alpha}\sigma_{\beta}^{2}+\sigma_{\alpha}^{2}G_{\beta}\right)^{2}\left(\sigma_{\alpha}+\sigma_{\beta}\right)/\lambda_{r}}{{\rm sinh}\left(\frac{d}{\lambda_{r}}\right)\prod_{\eta=\pm}\left(\mathcal{G}_{\alpha r}^{\left(\eta\right)}\sigma_{\beta}^{2}+\sigma_{\alpha}^{2}\mathcal{G}_{\beta r}^{\left(\eta\right)}\right)}
\end{equation}
reduces to Eq.~(\ref{part_hole_g}) at zero field,
while 
\begin{equation}
\mathcal{G}_{B} = \frac{1}{2}\left(\sigma_{\beta}^{2}-\sigma_{\alpha}^{2}\right)\sum_{\eta=\pm}\eta\frac{\mathcal{G}_{\beta r}^{\left(\eta\right)}G_{\alpha}-\mathcal{G}_{\alpha r}^{\left(\eta\right)}G_{\beta}}{\mathcal{G}_{\alpha r}^{\left(\eta\right)}\sigma_{\beta}^{2}+\sigma_{\alpha}^{2}\mathcal{G}_{\beta r}^{\left(\eta\right)}}
\end{equation}
is nonzero only when the magnetic field is applied; 
here $\mathcal{G}_{ir}^{(-)}\equiv G_{i}+\left({\sigma_{i}}/{\lambda_{r}}\right)\mathrm{Tanh}\left[{d}/{2\lambda_{r}}\right]$ while $\mathcal{G}_{ir}^{(+)}\equiv G_{i}+\left({\sigma_{i}}/{\lambda_{r}}\right)\mathrm{Coth}\left[{d}/{2\lambda_{r}}\right]$; the decay length $\lambda_r$ is given by the limit of $\lambda_1$ in the large $g$ limit, yielding $\lambda_r^{-2}=(r_\alpha/\sigma_\alpha+r_\beta/\sigma_\beta)/2$. Thus, we find that strong interactions do not radically alter the structural spin conductance in the sense that the spin conductance neither vanishes or diverges in this regime. When $d \gg \lambda_r$, Eq.~(\ref{large_g_g}) simplifies to:
 \begin{equation}
 \label{long_d_big_g}
\mathcal{G} =2   \frac{\left(\sigma_{\alpha}+\sigma_{\beta}\right)}{\lambda_{r}}\frac{\left(G_{\alpha}\sigma_{\beta}^{2}+\sigma_{\alpha}^{2}G_{\beta}\right)^{2}}{\left(\mathcal{G}_{\alpha r}\sigma_{\beta}^{2}+\sigma_{\alpha}^{2}\mathcal{G}_{\beta r}\right)^2}e^{-d/\lambda_r}\,.
 \end{equation}
 Thus we find that for large inter-band scattering, the nonlocal signal does not depend on $g$ but only on the decay processes (e.g. Gilbert damping) via $r_i$.

Third, we consider a finite magnetic field and the limit when magnons are non-interacting. In the zero coupling regime, $g=0$, one finds that the structural conductance is the sum of the parallel channels, $\mathcal{G}=\mathcal{G}_\alpha+\mathcal{G}_\beta$.  Here,
\begin{equation}
\label{noninteracting}
\mathcal{G}_i=\frac{(\sigma_i/\lambda_{ir}) G_i^{2}}{\left[\sigma^2_{i}/\lambda_{ir}^2+G_i^{2}\right]\mathrm{sinh}\left(\frac{d}{\lambda_{ir}}\right)+{2(\sigma_i/\lambda_{ir}) G_i}\mathrm{cosh}\left(\frac{d}{\lambda_{ir}}\right)}\,.
\end{equation}
where $\lambda_{ir} ^{-2}= r_i/\sigma_i$ is determined by decay processes. For $d \gg \lambda_{ir} $, we find that 
\begin{equation}
\label{noninteracting_long_d}
\mathcal{G}_i=\frac{2(\sigma_i/\lambda_{ir}) G_i^{2}}{({(\sigma_i/\lambda_{ir}) +G_i})^2}e^{-d/\lambda_{ir}}
\end{equation}
which shows an exponential decay over distance. 
When $\alpha$- and $\beta$-magnons are identical at the bulk and interfaces, both Eqs.~(\ref{noninteracting_long_d}) and (\ref{long_d_big_g}) reduce to Eq.~(\ref{part_hole_g}).

In the following sections we calculate and estimate the various parameters that enter into the phenomenological theory above. 

\section{Transport coefficients: Microscopic Theory}\label{sec:microscopictheory}
\begin{figure}[ht]
\includegraphics[width=\linewidth]{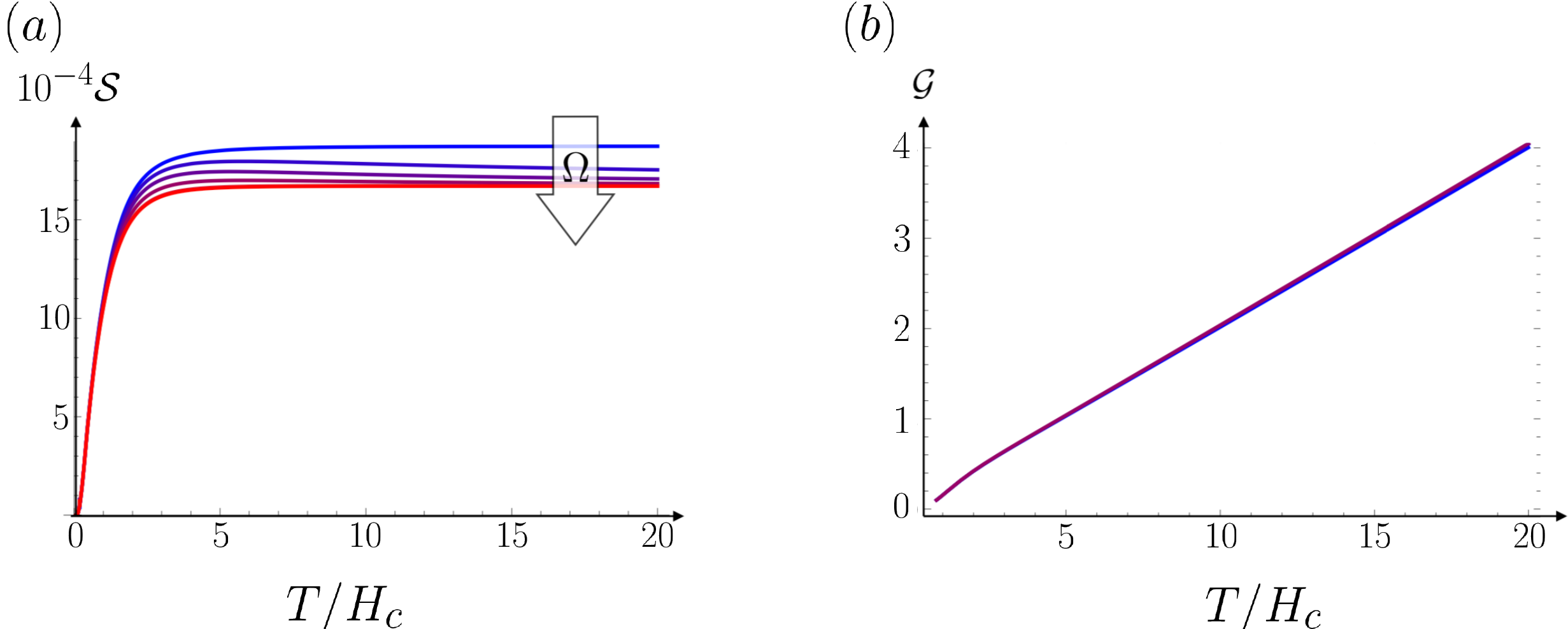}
\caption{(a) Structural Seebeck coefficient $\mathcal{S}$ and (b) structural spin conductance $\mathcal{G}$ as functions of temperature $T/H_c$ for a field $H= 0.2 H_c$. The temperature dependence of the inter-magnon scattering is given by $g = \Omega (T/T_c)^3$ (see Appendix \ref{app:scattlength}). Shown for both plots are $\Omega = 0, 1, 10, 10^3, 10^5$, corresponding to a shift from blue to red coloring. While increased scattering slightly diminishes the SSE, it has no discernible effect on the spin conductance for these particular parameters. For these plots, the parameters $g_a^{\uparrow\downarrow}=1/100\mathfrak{a}^2$ (which for $\mathfrak{a}=1\mathrm{\r{A}}$ corresponds to $g^{\uparrow \downarrow}\sim 1/\mathrm{nm}^2$), $k_F=1/\mathfrak{a}$, $6Js^2=2 H_c$, $\alpha=10^{-3}$ and $d=100\mathfrak{a}$ were used.}
\label{fig:structuralSeebeck}
\end{figure}

In this section, we compute the interfacial spin conductances from a microscopic model for the interface. In addition, the bulk magnon conductance, as well as the bulk Seebeck coefficient, are obtained in linear response from transport kinetic theory. Based on these results the structural Seebeck coefficient is evaluated and plotted in Figs. \ref{fig:structuralSeebeck}.  

\subsection{Contact magnon spin conductance}\label{sec:contact-magnon-conduc}
In this section, we compute interface transport coefficients appropriate to our bulk drift-diffusion theory above, allowing for the boundary conditions, Eqs~(\ref{bcL}) to be computed.

\begin{figure}[h!]
	\includegraphics[width=\linewidth,clip=]{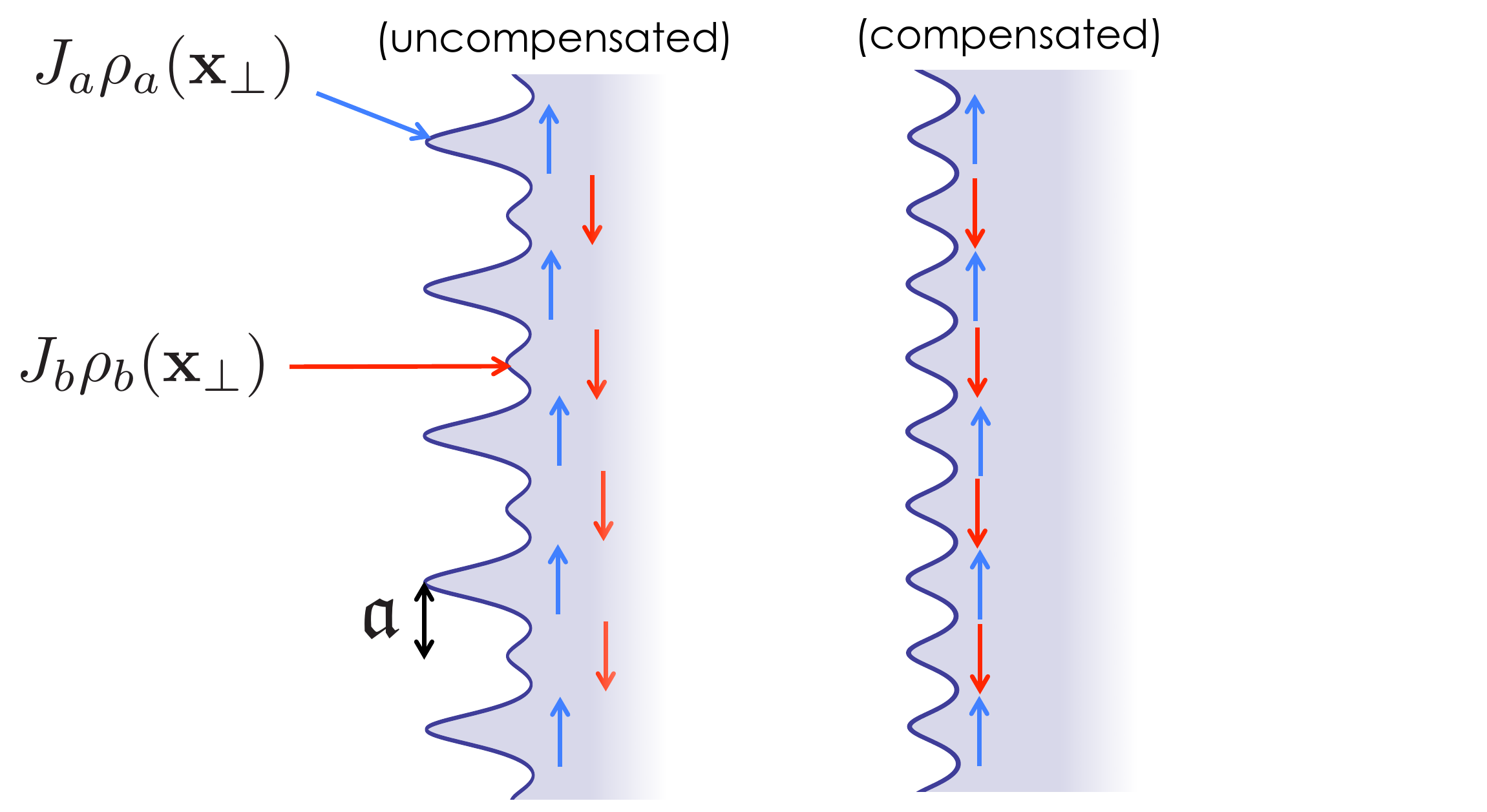}
	\caption{Effective spin densities of AF$|$NM interface as experienced by normal metal electrons scattering off of the interface, for the compensated and uncompensated cases.}
	\label{intsch}
\end{figure}

Let us suppose that the spin degrees of freedom of the AF are coupled to those of the normal metals by the exchange Hamiltonian (\ref{eq:contactHamiltonian}). Here it is understood that $\mathbf{i}$ labels the lattice along the interface (see Fig. \ref{intsch}). Specifically, the lattice is the set of vectors 
$\mathbf{R}_{2}=\left\{ n\mathfrak{a}\hat{\mathbf{z}}+m\mathfrak{a}\hat{\mathbf{y}}\right\}$. The integers  $n$ and $m$ are such that $\mathbf{i}$ corresponds to $a$- and $b$-atoms when $n+m$ are even and odd, respectively. In this model, we assume that $a$ and $b$ atoms are evenly spaced, which is not essential in what follows. Besides, the itinerant electronic density, corresponding to evanescent modes in the x-direction, decays over an atomistic distance inside the AF. The spin density of itinerant electrons in the normal metal is $\hat{\mathbf{S}}\left(\mathbf{x}\right)=\left(\hbar/2\right)\sum_{\sigma\sigma'}\hat{\Psi}_{\sigma}^{\dagger}\left(\mathbf{x}\right)\boldsymbol{\tau}_{\sigma\sigma'}\hat{\Psi}_{\sigma'}\left(\mathbf{x}\right)$, where $\hat{\Psi}_{\sigma}({\bf x})$ is the electron operator and $\boldsymbol{\tau}$ the Pauli matrix vector. The exchange coupling $J_{\mathbf{i}}=J_{a}$, if $\mathbf{i}\in a$, and $J_{\mathbf{i}}=J_{b}$, if $\mathbf{i}\in b$, while the local spin density at each lattice site $\mathbf{i}$ is modulated by the function $\rho_{\bf i}({\bf x})=\left|\phi_{\mathbf{i}}\left(\mathbf{x}\right)\right|^{2}$, with $\phi_{\mathbf{i}}$ the localized orbital at position $\mathbf{i}$ . Note that in general the orbitals for the $a$ and $b$ sublattices may be different. 

Based on the model represented by the contact Hamiltonian (\ref{eq:contactHamiltonian}), we wish to obtain the magnonic spin current across the interface using Fermi's Golden Rule. We expand $\hat{H}_{I}$ in terms of magnon operators up to order $n_i/s$, obtaining $\hat{H}_{I}=\hat{H}_{I}^{\left(\parallel\right)}+\hat{H}_{I}^{\left(\rm{sf}\right)}$. The first term is the coherent Hamiltonian $\hat{H}_{I}^{\left(\parallel\right)}=\sum_{\bf kk'} U_{\mathbf{k}\mathbf{k}'}\left(\hat{c}_{\mathbf{k}\uparrow}^{\dagger}\hat{c}_{\mathbf{k}'\uparrow}-\hat{c}_{\mathbf{k}\downarrow}^{\dagger}\hat{c}_{\mathbf{k}'\downarrow}\right)$,
with $\hat{c}_{\mathbf{k}\sigma}\left(\hat{c}^{\dagger}_{\mathbf{k}\sigma}\right)$ the fermionic operator that annihilate (create) and electron with spin-$\sigma$ and momentum $\mathbf{k}$. The term $\hat{H}_{I}^{\left(\parallel\right)}$ gives rise to coherent spin torques, and magnonic corrections to it, $\sim \mathbf{n}\times \boldsymbol{\mu}$. Since we are assuming a fixed order parameter $\mathbf{n}$ and focus only on thermal magnon spin currents, we need not consider this term. The second contribution, $\hat{H}_{I}^{\left(\rm{sf}\right)}$, is the spin-flip Hamiltonian that describes processes in which both branches of magnons are annihilated and created at the interface by spin-flip scattering of electrons. This term reads,
\begin{equation}
\hat{H}_{I}^{\left(\text{sf}\right)}=\sum_{\mathbf{q}\mathbf{k}\mathbf{k}'}\left(V^{\alpha}_{\mathbf{q}\mathbf{k}\mathbf{k}'}\hat{\alpha}_{\mathbf{q}}^{\dagger}\hat{c}_{\mathbf{k}\downarrow}^{\dagger}\hat{c}_{\mathbf{k}'\uparrow}+V^{\beta}_{\mathbf{q}\mathbf{k}\mathbf{k}'}\hat{\beta}_{\mathbf{q}}^{\dagger}\hat{c}_{\mathbf{k}\uparrow}^{\dagger}\hat{c}_{\mathbf{k}'\downarrow}\right)+h.c.,
\end{equation}
where the matrix elements are
\begin{align}
\nonumber	
V^{\alpha}_{\mathbf{q}\mathbf{k}\mathbf{k}'}&\equiv-\sqrt{\frac{8S}{N}}\int d{\bf x}\Psi_{\mathbf{k}}^{*}\left(\mathbf{x}\right)\Psi_{\mathbf{k}'}\left(\mathbf{x}\right)\\
&\times\left(\rho_{a}^{*}\left(\mathbf{q},\mathbf{x}\right)J_{a}\mathrm{cosh}\theta_{\mathbf{q}}-\rho_{b}^{*}\left(\mathbf{q},\mathbf{x}\right)J_{b}\mathrm{sinh}\theta_{\mathbf{q}}\right)
\end{align}
and
\begin{align}
\nonumber
V^{\beta}_{\mathbf{q}\mathbf{k}\mathbf{k}'}&\equiv-\sqrt{\frac{8S}{N}}\int d{\bf x}\Psi_{\mathbf{k}}^{*}\left(\mathbf{x}\right)\Psi_{\mathbf{k}'}\left(\mathbf{x}\right)\\
&\times\left(\rho_{b}\left(\mathbf{q},\mathbf{x}\right)J_{b}\mathrm{cosh}\theta_{\mathbf{q}}-\rho_{a}\left(\mathbf{q},\mathbf{x}\right)J_{a}\mathrm{sinh}\theta_{\mathbf{q}}\right).
\end{align}
Here, the function $\Psi_{\bf k}({\bf x})$ represent the eigenstates of the nonmagnetic Hamiltonian. Specifically,  in the yz directions, the wavefunction is a delocalized Bloch state of the interfacial lattice, which we assume here for simplicity to be common to the both the metal and insulators (as is common in such heterostuctures); in the x direction, the state is an evanescent mode on the insulator side of the interface, and a Bloch-like state of the metallic lattice on the other, which reduces to the usual 3D metallic Bloch state far inside the metal. The quantities $\rho_{a}$ and $\rho_b$ are defined by $\rho_{a}\left(\mathbf{q},\mathbf{x}\right)=\sum_{\mathbf{i}\in a}\rho_{\mathbf{i}}\left(\mathbf{x}\right)e^{i\mathbf{q}\cdot\mathbf{i}}$ and $\rho_{b}\left(\mathbf{q},\mathbf{x}\right)=\sum_{\mathbf{i}\in b}\rho_{\mathbf{i}}\left(\mathbf{x}\right)e^{i\mathbf{q}\cdot\mathbf{i}}$, with $\rho_{\mathbf{i}}(\mathbf{x})=\left|\phi_\mathbf{i}(\mathbf{x}) \right|^2$ as the density of the localized AF electron orbital at site $\mathbf{i}$. The quantities $\mathrm{cosh}\theta_{\mathbf{q}}$ and $\mathrm{sinh}\theta_{\mathbf{q}}$ originate from the Bogoliubov transformation that diagonalizes the noninteracting magnon Hamiltonian \cite{BogoParamet}.

It is instructive to consider the simplest case of interfacial spin transport. This occurs when the interface is fully compensated, i.e., $\rho_{i\in a}(\mathbf{x})=\rho_{i\in b}({\bf x}+\mathbf{i}))$ and $J_a=J_b$, see right side of Fig. \ref{intsch}. Because the normal metal electronic states $\Psi_\mathbf{k}(\mathbf{x})$ are Bloch states of the interfacial nonmagnetic Hamiltonian, then translation by the lattice spacing $\mathfrak{a}$ in the $y$ (or $z$) direction transform, $\Psi_\mathbf{k}\rightarrow e^{ik_y \mathfrak{a}}\Psi_\mathbf{k}$. Using $\rho_{a(b)}(\mathbf{q},\mathbf{x})=e^{i q_y\mathfrak{a}}\rho_{b(a)}(\mathbf{q},\mathbf{x}+\mathfrak{a}\hat{\mathbf{y}})=e^{i 2 q_y\mathfrak{a}}\rho_{a(b)}(\mathbf{q},\mathbf{x}+2\mathfrak{a}\hat{\mathbf{y}})$, it follows that $V^{\alpha}_{\mathbf{q}\mathbf{k}\mathbf{k}'}=e^{i2\mathfrak{a}\left(q+k-k' \right)_y}V^{\alpha}_{\mathbf{q}\mathbf{k}\mathbf{k}'}$. Applying translational invariance on the full Hamiltonian, under $\mathbf{x}\rightarrow\mathbf{x}+\mathfrak{a}\hat{\mathbf{y}}$, one has that this is independent of $\mathbf{q}+\mathbf{k}-\mathbf{k}'$, and it follows that $V^{\alpha}_{\mathbf{q}\mathbf{k}\mathbf{k}'}=e^{i\phi} \left(V^{\beta}_{\mathbf{q}\mathbf{k}'\mathbf{k}}\right)^{*}$, with the phase factor defined by $\phi=\mathfrak{a}\left(q_y+k_y-k'_y \right)$. Since all of the interfacial transport coefficients are proportional to $\left|V \right|^2$, we establish that they become identical for both magnonic branches, at zero field, for the case of fully compensated interface. 

It is also interesting to note the role played by Umklapp scattering processes at the interface.  Suppose again a fully compensated interface.  Then, in the small $\mathbf{q}$ limit, one finds $\mathrm{cosh}\theta_{\mathbf{q}}\approx \mathrm{sinh}\theta_{\mathbf{q}}$, and the matrix elements $V^{\alpha}_{\mathbf{q}\mathbf{k}\mathbf{k}'}$ and $V^{\beta}_{{q}\mathbf{k}\mathbf{k}'}$ vanish when $(\mathbf{q}-\mathbf{k}+\mathbf{k}')_\perp=0$ (specular scattering of electrons), where the subscript ``$\perp$" designates the in-plane components. However, when $(\mathbf{q}-\mathbf{k}+\mathbf{k}')_\perp=\mathbf{G}_{mn}$, where $\mathbf{G}_{mn}=n(\pi/\mathfrak{a})\hat{\mathbf{y}}+m(\pi/\mathfrak{a})\hat{\mathbf{z}}$ is the reciprocal lattice vector, the matrix elements do not vanish for odd values of $m+n$, and transport for each species becomes possible.  We therefore expect Umklapp scattering processes to play a crucial role in the low temperature behavior of the magnon conductance, as well as other interfacial linear transport coefficients.  This is consistent with Takei $et~al.$ \cite{TakeiPRB2014,chengprl2014}, where Umklapp scattering is found to be responsible for a finite spin-mixing conductance, describing coherent spin torques, at an AF$|$NM interface. Umklapp processes, however, may only happen when part of the $yz$ cross section of the normal metal $\left| \mathbf{k}\right|=2k_F$ surface lies outside the magnetic Brillioun zone of the lattice interface, for instance in a spherical Fermi surface this conditions becomes $2k_F>\pi/\mathfrak{a}$.

We return to the general case in order to obtain the contact magnon conductances $G_{\alpha,\beta}$. This can be done by a straightforward application of Fermi's Golden rule to calculate the magnonic spin current flowing across the interface. The magnon current is expressed as\cite{BenderPRB2015}
\begin{equation}
j_i=2D_F^{2}\int d\epsilon g^{(i)}_{\epsilon}\left|V_{i}\left(\epsilon\right)\right|^{2}(\epsilon-\mu'_i) \left[f_{im}(\epsilon)-f_{ie}(\epsilon)\right],
\label{jio}
\end{equation}
and therefore, the magnon spin current through the interface becomes $j_s=\hbar(j_\alpha-j_\beta)$. Here we have defined for $i=\alpha,\beta$, 
\begin{align}
\nonumber
\left|V_{i}\left(\epsilon\right)\right|^{2}=\frac{\pi d_{AF}}{D_F^{2}}\sum_{\mathbf{q}\mathbf{k}\mathbf{k}'}&\frac{1}{g^{(i)}_{\epsilon}}\left|V^i_{\mathbf{q}\mathbf{k}\mathbf{k}'}\right|^{2}\delta\left(\epsilon_{\mathbf{k}}-\epsilon_{F}\right)\\
&\quad\times \delta\left(\epsilon_{\mathbf{k}'}-\epsilon_{F}\right)\delta\left(\epsilon-\epsilon_{\mathbf{q}}\right),
\end{align}
with $D_F$ as the normal metal density of states and $g^{(i)}_{\epsilon}$ the $i$-magnon density of states. In Eq. (\ref{jio}) $\mu'_{\alpha,\beta}=\mp\mu$, where we recall that $\mu$ is the spin accumulation. The Bose-Einstein distribution for the i-magnons is $f_{im}(\epsilon)=1/[e^{( \epsilon-\mu_i)/k_BT_i}-1]$, and $f_{ie}(\epsilon)=1/[e^{( \epsilon -\mu_i')/k_BT_e}-1]$ corresponds to the effective electron-hole-excitation density experienced by the $i$-magnons.
\begin{figure}[h!]
	\includegraphics[width=\linewidth,clip=]{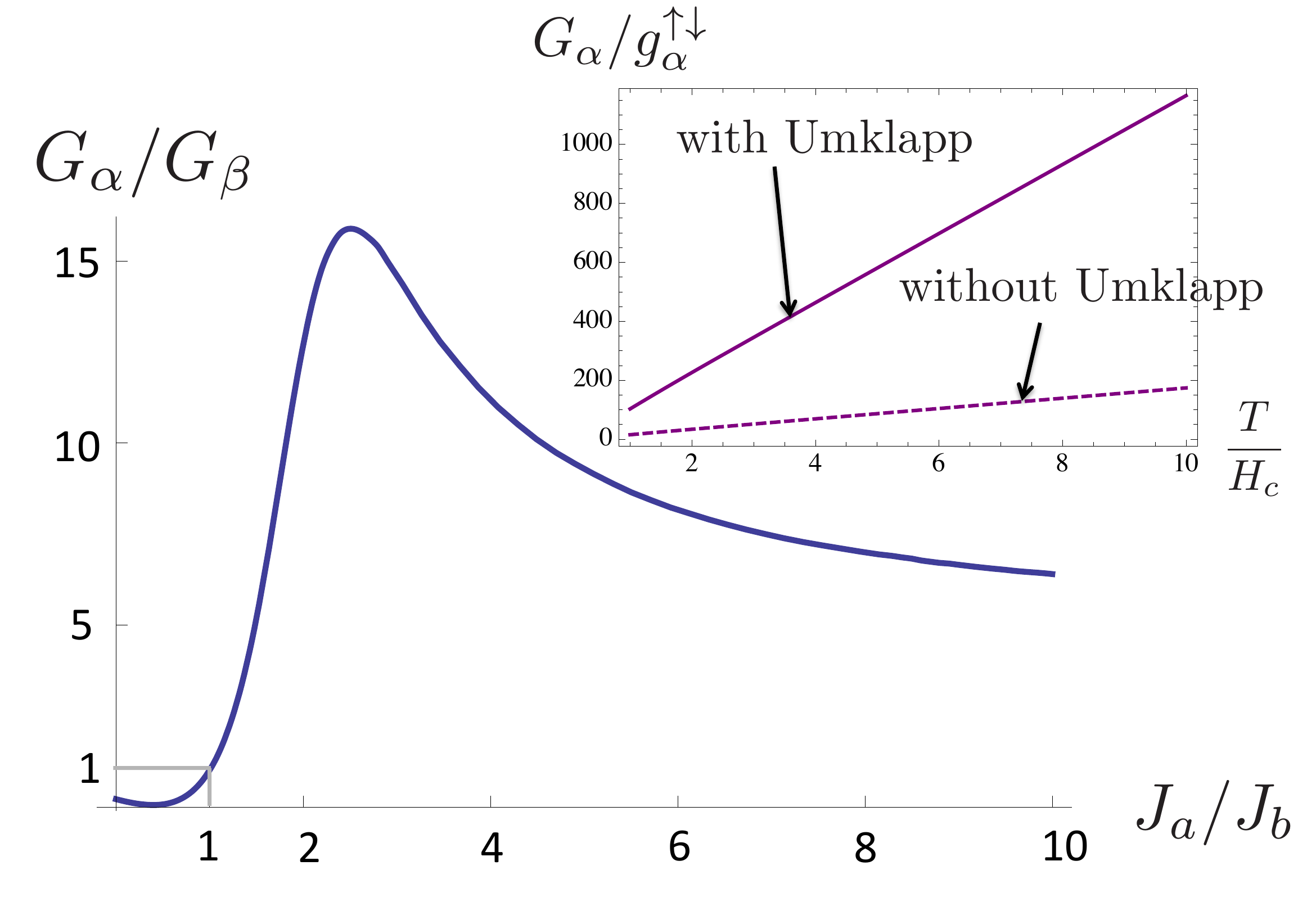}
	\caption{Ratio of interfacial conductances of the two magnon branches $\alpha$ and $\beta$ at zero field for different ratios of interfacial sublattice exchange constants.  Sublattice symmetry breaking ($J_a\neq J_b$) is necessary to obtain a structural spin Seebeck effect in the absence of magnetic fields (see Eq. (\ref{SbulkSymm})).  Inset: temperature dependence of $G_\alpha$ including and excluding Umklapp scattering ($m\neq0$ and/or $n\neq0$ in Eq.~(\ref{xis})).  All curves are obtained for $k_F=4/\mathfrak{a}$, and $6Js^2=2 H_c$.}
	\label{fig:GvJ}
\end{figure}

In a simple model, we may take the atomic densities for both sublattices as $\rho_i(\mathbf{x})=\delta(\mathbf{x}-\mathbf{r}_i)$ and the normal metal wavefunctions $\Psi_\mathbf{k}(\mathbf{x})=e^{i\mathbf{k}_\perp\cdot \mathbf{x}} F_\mathbf{k}(x)/\sqrt{\mathcal{V}_{NM}}$. Here the function $F_\mathbf{k}(x)$ describes the decay within the AF and ${\cal V}_{NM}$ the normal metal volume. Then, defining the spin-mixing conductance $g_{i}^{\uparrow\downarrow}\equiv16\pi N d_{AF}\mathcal{V}_{AF}\left(s\eta D_{F}J_{i}/\mathcal{V}_{NM}\right)^{2}$, where $\eta=\int dx F_\mathbf{k'}(x)F^*_\mathbf{k}(x)$, which we take to be momentum independent for simplicity, one may write the $i$-magnon spin current $\hbar j_{i}$ in Eq. (\ref{jio}) as
\begin{equation}
\hbar j_{i}=\frac{1}{16\pi}\left(g_{a}^{\uparrow\downarrow}\xi_{aa}^{\left(i\right)}+g_{b}^{\uparrow\downarrow}\xi_{bb}^{\left(i\right)}+2\sqrt{g_{a}^{\uparrow\downarrow}g_{b}^{\uparrow\downarrow}}\xi_{ab}^{\left(i\right)}\right),
\label{ji}
\end{equation}
where the functions $\xi^{(i)}_{ll'}$, which carry units of energy, are given by
\begin{align}
\nonumber
\xi^{(i)}_{ll'}&=\frac{1}{D_F^2 (s \mathcal{V}_{AF} )}\sum_{\mathbf{q},\mathbf{k},\mathbf{k}'}\sum_{m,n}\mathcal{F}^{(ll')}_{mn\mathbf{q}}\left(\epsilon_\mathbf{q}- \mu_i\right)\left(n_{im}-n_{ie}\right)\\
&\qquad\qquad\times\delta(\epsilon_F-\epsilon_\mathbf{k})\delta(\epsilon_F-\epsilon_{\mathbf{k}'})\delta_{\mathbf{k'}-\mathbf{k}-\mathbf{q},\mathbf{G}_{mn}},
\label{xis}
\end{align}
with $\mathcal{F}^{(aa)}_{mn\mathbf{q}}=\rm{cosh}^2\theta_\mathbf{q}$, $\mathcal{F}^{(bb)}_{mn\mathbf{q}}=\rm{sinh}^2\theta_\mathbf{q}$ and $\mathcal{F}^{(ab)}_{mn\mathbf{q}}=\mathcal{F}^{(ba)}_{mn\mathbf{q}}=(-1)^{m+n+1}\rm{cosh}\theta_\mathbf{q}\rm{sinh}\theta_\mathbf{q}$. The motivation for expressing the spin current in the form of Eq. (\ref{ji}) is that in the case of particle-hole symmetry at the interface, $\hbar j_i=(g^{\uparrow\downarrow}/4\pi)\xi\sim (g^{\uparrow\downarrow}/4\pi)(\hbar\omega-\mu_i) (n_i/s)$. In particular, the $m=n=0$ term in Eq. (\ref{xis}) gives spectral scattering processes, while all others ($m\neq n\neq 0$) correspond to Umklapp scattering. 

The contact spin conductances $G_{\alpha}$ and $G_{\beta}$ are obtained by the linearization of the $i$-magnon current given by Eq. (\ref{ji}), i.e., $G_i=\left. \left(\partial j_i/\partial{\mu_i}\right)\right|_{\mu_i=0}$. The ratio of interfacial conductances of the two magnon branches $\alpha$ and $\beta$ is shown in Fig. \ref{fig:GvJ} at zero field and as a function of ratios of interfacial sublattice exchange constants $J_{a}/J_{b}$. The ratio $G_{\alpha}/G_{\beta}$ reaches a maximum value to later saturates when $J_{a}/J_{b}$ is increased. In particular, we note that $G_{\alpha}=G_{\beta}$ when the interfacial exchange constants are equal. Thus, the breaking of sublattice symmetry ($J_a\neq J_b$) is necessary in realizing a structural spin Seebeck effect in the absence of a field, as is seen from Eq. (\ref{SbulkSymm}). In the inset of Fig. \ref{fig:GvJ} we display the temperature dependence of $G_{\alpha}$. In this plot we have included (solid line) and excluded (dashed line) Umklapp scattering.

\begin{figure}[h!]
\includegraphics[width=1\linewidth,clip=]{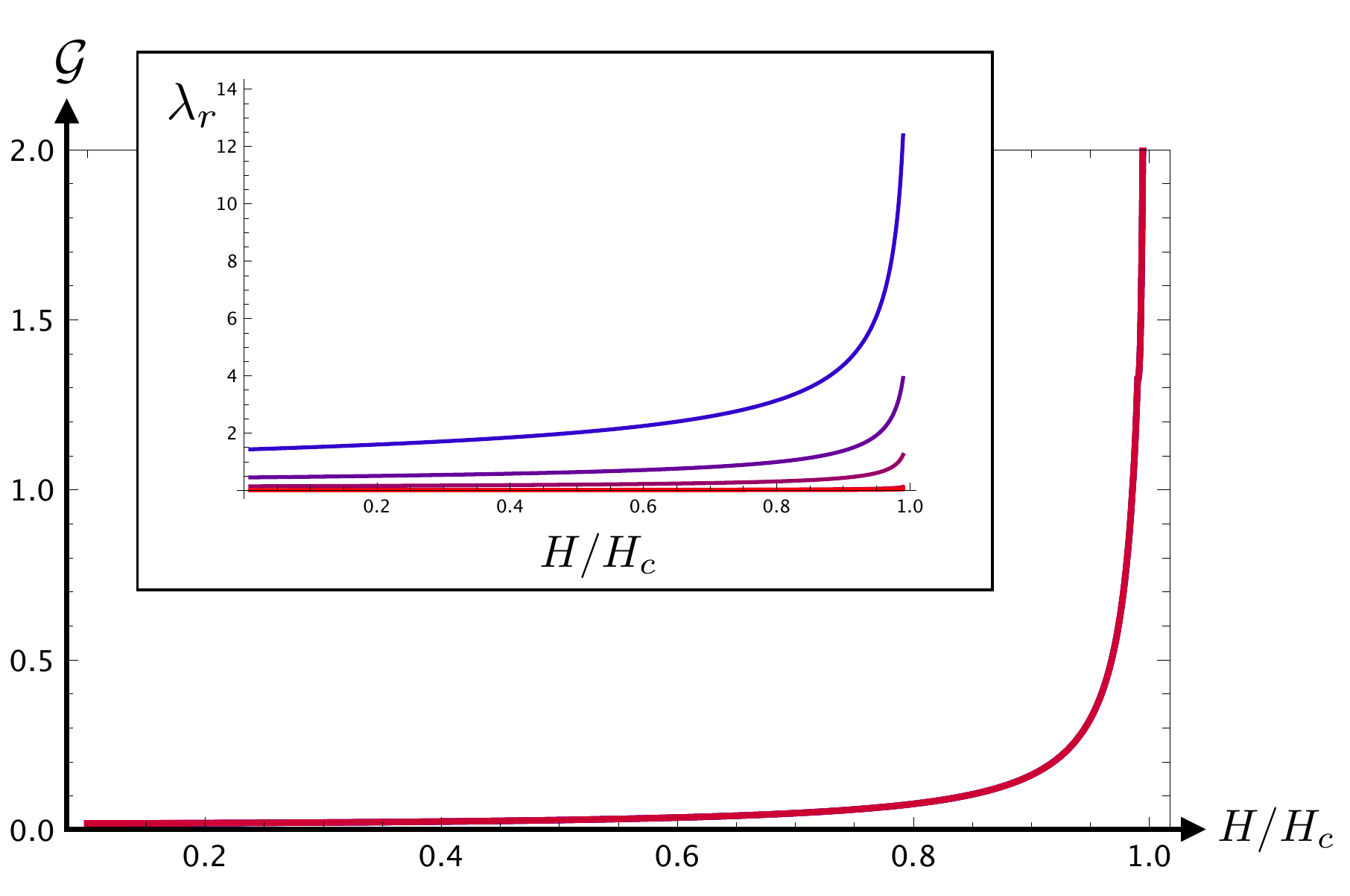}
\caption{Main figure: behavior of conductance $\mathcal{G}$ near spin flop. While the spin diffusion length $\lambda_r$ diverges as $|H|\rightarrow H_c$, the conductance $\mathcal{G}$, though sharply increasing, does not actually diverge because of bottlenecking by the interface impedances; for noninteracting magnons it has a maximium value of $\max(G_\alpha/2,G_\beta/2)$ (see Eq.~(\ref{spin_flop_noninter})). The colors and parameters are identical to those shown in Fig.~\ref{fig:structuralSeebeck}.}
\label{SpinFlop}
\end{figure}

\subsection{Bulk magnon conductances and spin Seebeck coefficients}\label{subsec:bulkconduc-seebeck}
In this section, we evaluate the bulk magnon conductances $\sigma_{\alpha,\beta}$ and bulk spin Seebeck coefficients $\varsigma_{\alpha,\beta}$. These are obtained from standard kinetic theory of transport. Unlike previous works \cite{RezendePRB2016,RezendePRB2016b,OhnumaPRB2013}, here we consider the magnonic transport driven, in addition to thermal gradients, by spin biasing. The generic expressions for the magnon current in the bulk are,
\begin{align}\label{currentcoefs}
j_{i}=-\int\frac{d{\bf q}}{\left(2\pi\right)^{3}}\tau_{i}v_{i}^{2}\frac{\partial f_{i}}{\partial x},
\end{align}
where the integration is over the Brillouin zone and $i=\alpha,\beta$. The magnon relaxation time is $\tau_{i}$ and the magnon group velocity along the $x$ direction is $v_{i}=\partial\epsilon_{i}/\hbar\partial{k_{x}}$. The number of i-magnons with momentum ${\bf q}$ is denoted by $f_i$ and given by the Bose-Einstein distribution function. This yields the transport coefficients,
\begin{align}
\sigma_i&\label{eq:sigma}=\frac{4\tilde{J}^{4}H^2_c}{9\hbar^2}\int\frac{d{\bf q}}{(2\pi)^{3}}\frac{{\tau}_i\sin^2(a{q}_x) \gamma_\mathbf{q}^2}{1+\tilde{J}^{2}\left(1-\gamma_\mathbf{q}^2\right)}\frac{{\beta}e^{\beta\epsilon_i}}{\left(e^{\beta\epsilon_i}-1\right)^{2}},\\
\varsigma_i&\label{eq:varsigma}=\frac{4\tilde{J}^{4}H^2_c}{9\hbar^2}\int\frac{d{\bf q}}{(2\pi)^{3}}\frac{{\tau}_i\sin^2(a{q}_x) \gamma_\mathbf{q}^2}{1+\tilde{J}^{2}\left(1-\gamma_\mathbf{q}^2\right)}\frac{{\epsilon}_i{\beta}^2e^{\beta\epsilon_i}}{\left(e^{\beta\epsilon_i}-1\right)^{2}},
\end{align} 
where $\tilde{J}\equiv6Js^2/H_c$ is roughly the N\'eel temperature in units of $H_c$ and $\beta=1/k_BT$. Similarly, we may obtain expressions for the damping rates $r_i$ from $\left.\dot{n}_i\right|_{\alpha}=\left(2\pi\right)^{-3}\int{d{\bf q}}\,{n_i}/{\tau_{i\mathrm{g}}}$ with $\tau_{i\mathrm{g}}$ the Gilbert damping lifetime. From the above relation $r_{i}$ is extracted and obeys
\begin{equation}
r_i=\frac{2\alpha}{\hbar}\int\frac{d{\bf q}}{\left(2\pi\right)^{3}}\frac{\beta \epsilon^2_i e^{\beta\epsilon_i}}{\left(e^{\beta\epsilon_i}-1\right)^2},
\end{equation}
and $\alpha$ is being the Gilbert damping constant. 

The momentum relaxation rate, entering in the transport coefficients obtained in Eqs. (\ref{eq:sigma}) and (\ref{eq:varsigma}), has contributions from different sources; Gilbert damping, disorder scattering, magnon-phonon scattering and $\alpha$-magnon--$\beta$-magnon scattering. For simplicity, we consider the regime in which Gilbert damping dominates transport:
\begin{equation}
\tilde{\tau}^{-1}_{i}\approx\tilde{\tau}^{-1}_{i\rm{g}},
\label{gilbert}
\end{equation}
where $\tilde{\tau}^{-1}_{i\rm{g}}=2\alpha{\epsilon}_i/H_c$. Note that tilde represents units of $H_c$.

In Figs. \ref{fig:structuralSeebeck}~(a) and (b) we show the temperature-dependence of the spin Seebeck coefficient and structural spin condutance, respectively using the interface and bulk transport coefficients above. The interaction parameter $g$ grows with temperature (see Appendix \ref{app:scattlength} and \ref{app:seebeckcoef}). As shown in Fig. \ref{fig:structuralSeebeck}~(a), however, the effects of $g$ are minimal, suppressing the spin Seebeck signal slightly and negligibly affecting the structural conductance.

\section{Spin-Flop Transition}
The spin-flop transition occurs as $|H| \rightarrow H_c$ from below. Here, the magnon spectrum becomes gapless and quadratic at low energies for one of the two magnon bands (say, the $\beta$-band, for purposes of discussion). When Gilbert damping dominates the transport time (Eq.~(\ref{gilbert})), the bulk conductance $\sigma_\beta$ in Eq.~(\ref{eq:sigma}) demonstrates an infrared divergence, while $\varsigma_\beta$, $r_\beta$ and $G_\beta$ are finite.  It is straightforward to show that the Seebeck coefficient, Eq.~(\ref{eq:fullstrucSeebeck}), does not diverge in this case, consistent with \cite{BenderPRL2017}.

In contrast to \cite{BenderPRL2017}, however, the structural conductance $\mathcal{G}$ does not diverge in the diffusive regime. Here, it is instructive to consider the noninteracting case, Eq.~(\ref{noninteracting}), which reduces to
\begin{equation}
\label{spin_flop_noninter}
\mathcal{G} = \frac{G_\beta^2}{(\sigma_\beta/\lambda_{\beta r})(d/\lambda_{\beta r})+2G_\beta}\,,
\end{equation}
which shows an \textit{algebraic}, rather than exponential, decay in film thickness, due to a diverging decay length $\lambda_{\beta r}$ ($\sim \sqrt{\sigma_\beta}$). For a thin film, this becomes $\mathcal{G}=G_\beta/2$; while the AF bulk shows zero spin resistivity ($\sigma_\beta^{-1} = 0$) due to the Bose-Einstein divergence at low energies, structural transport is bottlenecked by the interface resistance $G_\beta^{-1}$, which is only well defined in the diffusive regime. (The effect is similar to a superconducting circuit, which with perfectly conducting components, shows a finite resistance due to normal metal contacts.) While the signal does not diverge, there is a clear enhancement due to the diminished spin resistivity, as well as long-distance transport (algebraic in $d$), manifesting as a peak in the signal near the spin-flop transition \cite{Lebrun2018} (see Fig.~\ref{SpinFlop}). A full calculation for nonlocal spin injection - including spin Hall/inverse spin Hall effects absent here - would show additional impedances to spin flow due to spin resistance in the normal metal injector and detector.

\section{Conclusion and Discussion}\label{sec:con-disc}

In summary, we have presented a study of spin transport of magnons in insulating AFs in contact with normal metals. We focus on the thick-film limit, wherein a diffusive regime can be assumed and magnons are in a local thermodynamical equilibrium. The excitation of magnon currents is considered in linear response and driven by either a temperature gradient and/or spin biasing. The spin transport is studied by evaluating the structural magnon conductance and spin Seebeck effect within a phenomenological spin-diffusion transport theory. These parameters were calculated in terms of bulk transport coefficients as well as contact magnon conductances. While the former were computed through kinetic theory of transport, the latter are obtained from a microscopic model of the NM$|$AF interface. Furthermore, we allowed for the breaking of sublattice symmetry at the interface assuming an uncompensated magnetic order. In addition, the field- and temperature-dependence of the inter-magnon scattering rates, which redistribute angular momentum between the magnon branches, were estimated. We find that the effects of inter-magnon scattering, which lock the two magnon bands together, is negligible. Furthermore, we show that even as the bulk spin resistivity vanishes near the spin flop transition, normal metal|magnet interface spin impedance ultimately bottleneck transport, irrespective of interactions, in contrast to the stochastic theory \cite{BenderPRL2017} for thin films.

The phenomenological approach above ultimately breaks down for strong interactions (which occur near the spin-flop transition), where the individual $\alpha$ and $\beta$ clouds are no longer internally equilibrated with well-defined chemical potentials an interactions. Instead, a treatment of the interacting clouds (e.g. a kinetic theory approach) beyond the quasiequilibrium approach that is adopted here is needed. In addition, more sophisticated treatments of the transport time $\tau_i$ have been shown to more realistically reproduce experimental results \cite{RezendePRB2016}; such quasi-empirical transport times could be incorporated directly into our phenomenology.  Most importantly, our somewhat artificial assumption that the magnetic field is applied along the easy-axis in not necessarily realized in experiment. Instead, even simple bipartite AFs such as those modeled by our phenomenology above show complex paramagnetic behavior in response to a field applied along different axes. In these scenarios, heterostuctures may manifest both antiferromagnetic and ferromagnetic transport behaviors \cite{Lebrun2018}. Future work, such as the drift-diffusion approach discussed above, is needed to fully understand such scenarios at a more fundamental level. 

\begin{acknowledgments}
This work was supported by the European Union's Horizon 2020 Research and Innovation Programme under Grant DLV-737038 "TRANSPIRE", the Research Council of Norway through is Centres of Excellence funding scheme, Project No. 262633, "QuSpin" and European Research Council via Advanced Grant No. 669442 "Insulatronics". Also we acknowledge funding from the Stichting voor Fundamenteel Onderzoek der Materie (FOM) and the European Research Council via Consolidator Grant number 725509 SPINBEYOND.

Note added--  During the submission of our work, we became aware of another related article \cite{ShenPRB2019} that considers magnon transport in AFs.
\end{acknowledgments}

\appendix
\begin{widetext}
\section{Magnon-magnon interactions}\label{app:scattamp}
We start out by defining the AF Hamiltonian. Introducing a square lattice, labelling the sites in the lattice by ${\bs i}$, on sub-lattices $\cal A$ and $\cal B$, the nearest-neighbor Hamiltonian reads
\begin{equation}\label{eq:Haf}
\hat{H}_{AF}=J\sum_{\langle {\bs i}\in{\cal A},{\bs j}\in{\cal B}\rangle}\hat{\mathbf{s}}_{\mathbf{i}}\cdot\hat{\mathbf{s}}_{\mathbf{j}}-H\sum_{\bs i\in {\cal A},{\cal B}}\hat{s}_{\mathbf{i}z}-\frac{\kappa}{2s}\sum_{\bs i\in {\cal A},{\cal B}}\hat{\mathbf{s}}_{\mathbf{i}z}^{2},
\end{equation}
where $J>0$ is the exchange coupling, $H$ the magnetic field and $\kappa>0$ the uniaxial easy-axis anisotropy. Introducing the Holstein-Primakoff transformation, assuming a bipartite ground state, the spin operators in the limit of small spin fluctuations reads
\begin{subequations}
	\begin{align}
	\hat{\mathbf{s}}^{\cal A}_{\mathbf{i}z}&=s-a^{\dagger}_{\bs i}a_{\bs i},\qquad\qquad\quad\qquad\qquad \hat{\mathbf{s}}^{\cal B}_{\mathbf{i}z}=-s+b^{\dagger}_{\bs i}b_{\bs i},\\
	\hat{\mathbf{s}}^{\cal A}_{\mathbf{i}+}&=\sqrt{2s}a_{\bs i}-\frac{1}{\sqrt{2s}}a^{\dagger}_{\bs i}a_{\bs i}a_{\bs i},\qquad \qquad \hat{\mathbf{s}}^{\cal B}_{\mathbf{i}+}=\sqrt{2s}b^{\dagger}_{\bs i}-\frac{1}{\sqrt{2s}}b^{\dagger}_{\bs i}b^{\dagger}_{\bs i}b_{\bs i},\\
	\hat{\mathbf{s}}^{\cal A}_{\mathbf{i}-}&=\sqrt{2s}a^{\dagger}_{\bs i}-\frac{1}{\sqrt{2s}}a^{\dagger}_{\bs i}a^{\dagger}_{\bs i}a_{\bs i}\qquad\qquad \hat{\mathbf{s}}^{\cal B}_{\mathbf{i}-}=\sqrt{2s}b_{\bs i}-\frac{1}{\sqrt{2s}}b^{\dagger}_{\bs i}b_{\bs i}b_{\bs i}.
	\end{align}
\end{subequations}

The AF Hamiltonian Eq. (\ref{eq:Haf}) is expanded up to fourth order in the magnon operators, Fourier transformed through the relations $a_{\bs i}=\frac{1}{\sqrt{N}}\sum_{\bs i}e^{i{\bs k}\cdot{\bs i}}a_{\bs k}$ and $b_{\bs i}=\frac{1}{\sqrt{N}}\sum_{\bs i}e^{i{\bs k}\cdot{\bs j}}b_{\bs k}$
and expressed as ${H}_{AF}=E_0+H^{(2)}_{AF}+H^{(4)}_{AF}$ where
\begin{align}
\hat{H}^{(2)}_{AF}&\label{eq:H2AF}=(Jsz+\kappa)\sum_{\bf q}\left[(1+h)a^{\dagger}_{\bf q}a_{\bf q}+(1-h)b^{\dagger}_{\bf q}b_{\bf q}+\xi\gamma_{\bf q}(a_{\bf q}b_{-\bf q}+a^{\dagger}_{\bf q}b^{\dagger}_{-\bf q})\right]\\
\hat{H}^{(4)}_{AF}&=\nonumber-\frac{Jz}{2N}\sum_{{\bf q}_1{\bf q}_2{\bf q}_3{\bf q}_4}\delta_{{\bf q}_1+{\bf q}_2-{\bf q}_3-{\bf q}_4}\left[2\gamma_{{\bf q}_2-{\bf q}_4}a^{\dagger}_{\bf q_1}b^{\dagger}_{-\bf q_4}a_{\bf q_3}b_{-\bf q_2}+\frac{\kappa}{2s}\left(a^{\dagger}_{\bf q_1}a^{\dagger}_{\bf q_2}a_{\bf q_3}a_{\bf q_4}+b^{\dagger}_{\bf q_1}b^{\dagger}_{\bf q_2}b_{\bf q_3}b_{\bf q_4}\right)\right.\\
&\label{eq:H4AF}\left.\qquad\qquad\qquad\qquad\qquad\qquad+\gamma_{{\bf q}_4}\left(b^{\dagger}_{{\bf q}_1}b_{-{\bf q}_2}b_{{\bf q}_3}a_{{\bf q}_4}+b^{\dagger}_{{\bf q}_3}b^{\dagger}_{-{\bf q}_2}b_{{\bf q}_1}a^{\dagger}_{{\bf q}_4}+a^{\dagger}_{\bf q_1}a_{-\bf q_2}a_{\bf q_3}b_{{\bf q_4}}+a^{\dagger}_{\bf q_3}a^{\dagger}_{-\bf q_2}a_{\bf q_1}b^{\dagger}_{{\bf q_4}}\right)\right]
\end{align}
with $h=H/(Jsz+\kappa)$, $\xi=Jsz/(Jsz+\kappa)$ and $\gamma_{\bf q}=\frac{2}{z}\sum_{{\bs \delta}}\cos\left[{\bf q}\cdot{\bs \delta}\right]$ where $z$ is the coordination number. The quadratic part of the Hamiltonian, Eq. (\ref{eq:H2AF}), is diagonalized by the Bogoliubov transformation
\begin{align}
\hat{a}_{\bs q}&\label{eq: bogoliubov1}=l_{\bs q}\hat{\alpha}_{\bs q}+m_{\bs q}\hat{\beta}^{\dagger}_{-\bs q}\\
\hat{b}^{\dagger}_{-\bs q}&\label{eq: bogoliubov2}=m_{\bs q}\hat{\alpha}_{\bs q}+l_{\bs q}\hat{\beta}^{\dagger}_{-\bs q}
\end{align}
with the coefficients $l_{\bf q}=\left(\frac{(Jsz+\kappa)+\epsilon_{\bf q}}{2\epsilon_{\bf q}}\right)^{1/2}$, $m_{\bf q}=-\left(\frac{(Jsz+\kappa)-\epsilon_{\bf q}}{2\epsilon_{\bf q}}\right)^{1/2}\equiv-\chi_{\bf q}l_{\bf q}$ and $\epsilon_{\bf q}=(Jsz+\kappa)\sqrt{1-\xi^2\gamma^2_{\bf q}}$, resulting in Eq. (\ref{eq:diagHam}). In the diagonal basis, the interacting Hamiltonian Eq. (\ref{eq:H4AF}) finally becomes 
\begin{align}\label{eq:magmaghamiltonian}
\hat{H}^{(4)}_{AF}=\sum_{\mathbf{q}_1{\bf q}_2{\bf q}_3{\bf q_4}}&\nonumber\delta_{{\bf q}_1+{\bf q}_2-{\bf q}_3-{\bf q}_4}\left[V^{(1)}_{{\bf q}_1{\bf q}_2{\bf q}_3{\bf q}_4}\alpha^{\dagger}_{{\bf q}_1}\alpha^{\dagger}_{{\bf q}_2}\alpha_{{\bf q}_3}\alpha_{{\bf q}_4}+V^{(2)}_{{\bf q}_1{\bf q}_2{\bf q}_3{\bf q}_4}\alpha^{\dagger}_{{\bf q}_1}\beta_{-{\bf q}_2}\alpha_{{\bf q}_3}\alpha_{{\bf q}_4}+V^{(3)}_{{\bf q}_1{\bf q}_2{\bf q}_3{\bf q}_4}\alpha^{\dagger}_{{\bf q}_1}\alpha^{\dagger}_{{\bf q}_2}\alpha_{{\bf q}_3}\beta^{\dagger}_{-{\bf q}_4}\right.\\
&\nonumber\left.+V^{(4)}_{{\bf q}_1{\bf q}_2{\bf q}_3{\bf q}_4}\alpha^{\dagger}_{{\bf q}_1}\beta_{-{\bf q}_2}\alpha_{{\bf q}_3}\beta^{\dagger}_{-{\bf q}_4}+V^{(5)}_{{\bf q}_1{\bf q}_2{\bf q}_3{\bf q}_4}\beta_{-{\bf q}_1}\beta_{-{\bf q}_2}\alpha_{{\bf q}_3}\beta^{\dagger}_{-{\bf q}_4}+V^{(6)}_{{\bf q}_1{\bf q}_2{\bf q}_3{\bf q}_4}\alpha^{\dagger}_{{\bf q}_1}\beta_{-{\bf q}_2}\beta^{\dagger}_{-{\bf q}_3}\beta^{\dagger}_{-{\bf q}_4}\right.\\
&\left.+V^{(7)}_{{\bf q}_1{\bf q}_2{\bf q}_3{\bf q}_4}\alpha^{\dagger}_{{\bf q}_1}\alpha^{\dagger}_{{\bf q}_2}\beta^{\dagger}_{-{\bf q}_3}\beta^{\dagger}_{-{\bf q}_4}+V^{(8)}_{{\bf q}_1{\bf q}_2{\bf q}_3{\bf q}_4}\beta_{-{\bf q}_1}\beta_{-{\bf q}_2}\alpha_{{\bf q}_3}\alpha_{{\bf q}_4}+V^{(9)}_{{\bf q}_1{\bf q}_2{\bf q}_3{\bf q}_4}\beta_{-{\bf q}_1}\beta_{-{\bf q}_2}\beta^{\dagger}_{-{\bf q}_3}\beta^{\dagger}_{-{\bf q}_4}\right]
\end{align}
where the scattering amplitudes are $V^{(a)}_{{\bf q}_1{\bf q}_2{\bf q}_3{\bf q}_4}=-\left(\frac{Jz}{N}\right)l_{{\bf q}_1}l_{{\bf q}_2}l_{{\bf q}_3}l_{{\bf q}_4}\Phi^{(a)}_{{\bs 1}{\bs 2}{\bs 3}{\bs 4}}$. The functions $\Phi^{(a)}$ are the following expressions

\begin{align}
\Phi^{(1)}_{{\bs 1}{\bs 2}{\bs 3}{\bs 4}}&\label{eq:phi1}=\gamma_{{\bf q}_2-{\bf q}_4}\chi_{{\bf q}_2}\chi_{{\bf q}_4}-\frac{1}{2}\left(\gamma_{{\bf q}_2}\chi_{{\bf q}_2}+\gamma_{{\bf q}_4}\chi_{{\bf q}_4}+\gamma_{{\bf q}_2}\chi_{{\bf q}_1}\chi_{{{\bf q}_3}}\chi_{{{\bf q}_4}}+\gamma_{{\bf q}_4}\chi_{{{\bf q}_1}}\chi_{{{\bf q}_2}}\chi_{{{\bf q}_3}}\right)+\frac{\kappa}{2Jzs}\left(1+\chi_{{\bf q}_1}\chi_{{\bf q}_2}\chi_{{\bf q}_3}\chi_{{\bf q}_4}\right)
\\
\Phi^{(2)}_{{\bs 1}{\bs 2}{\bs 3}{\bs 4}}&\label{eq:phi2}\nonumber=-\gamma_{{\bf q}_2-{\bf q}_4}\chi_{{\bf q}_4}-\gamma_{{\bf q}_1-{\bf q}_4}\chi_{{\bf q}_1}\chi_{{\bf q}_2}\chi_{{\bf q}_4}+\gamma_{{\bf q}_4}\chi_{{{\bf q}_1}}\chi_{{{\bf q}_3}}+\gamma_{{\bf q}_4}\chi_{{{\bf q}_2}}\chi_{{{\bf q}_4}}+\frac{1}{2}\left(\chi_{{{\bf q}_3}}\chi_{{{\bf q}_4}}\left(\gamma_{{\bf q}_1}+\gamma_{{\bf q}_2}\chi_{{\bf q}_1}\chi_{{\bf q}_2}\right)+\left(\gamma_{{\bf q}_2}+\gamma_{{\bf q}_1}\chi_{{{\bf q}_1}}\chi_{{{\bf q}_2}}\right)\right)\\
&-\frac{\kappa}{Jzs}\left(\chi_{{\bf q}_2}+\chi_{{\bf q}_1}\chi_{{\bf q}_3}\chi_{{\bf q}_4}\right)
\\
\nonumber\Phi^{(3)}_{{\bs 1}{\bs 2}{\bs 3}{\bs 4}}&\label{eq:phi3}=-\gamma_{{\bf q}_2-{\bf q}_4}\chi_{{\bf q}_2}-\gamma_{{\bf q}_2-{\bf q}_3}\chi_{{\bf q}_2}\chi_{{\bf q}_3}\chi_{{\bf q}_4}+\gamma_{{\bf q}_2}\chi_{{\bf q}_1}\chi_{{\bf q}_3}+\gamma_{{\bf q}_2}\chi_{{\bf q}_2}\chi_{{\bf q}_4}+\frac{1}{2}\left(\chi_{{\bf q}_1}\chi_{{\bf q}_2}\left(\gamma_{{\bf q}_3}+\gamma_{{\bf q}_4}\chi_{{\bf q}_3}\chi_{{\bf q}_4}\right)+\left(\gamma_{{\bf q}_4}+\gamma_{{\bf q}_3}\chi_{{\bf q}_3}\chi_{{\bf q}_4}\right) \right)\\
&-\frac{\kappa}{Jzs}\left(\chi_{{\bf q}_4}+\chi_{{\bf q}_1}\chi_{{\bf q}_2}\chi_{{\bf q}_3}\right)
\\
\Phi^{(4)}_{{\bs 1}{\bs 2}{\bs 3}{\bs 4}}&\label{eq:phi4}\nonumber=\gamma_{{\bf q}_2-{\bf q}_4}+\gamma_{{\bf q}_1-{\bf q}_4}\chi_{{\bf q}_1}\chi_{{\bf q}_2}+\gamma_{{\bf q}_2-{\bf q}_3}\chi_{{\bf q}_3}\chi_{{\bf q}_4}+\gamma_{{\bf q}_1-{\bf q}_3}\chi_{{\bf q}_1}\chi_{{\bf q}_2}\chi_{{\bf q}_3}\chi_{{\bf q}_4}+\frac{2\kappa}{Jzs}\left(\chi_{{\bf q}_2}\chi_{{\bf q}_4}+\chi_{{\bf q}_1}\chi_{{\bf q}_3}\right)\\
&-\left(\chi_{{\bf q}_1}\left(\gamma_{{\bf q}_3}+\gamma_{{\bf q}_4}\chi_{{\bf q}_3}\chi_{{\bf q}_4}\right)+\chi_{{\bf q}_3}\left(\gamma_{{\bf q}_1}+\gamma_{{\bf q}_2}\chi_{{\bf q}_1}\chi_{{\bf q}_4}\right) +\chi_{{\bf q}_4}\left(\gamma_{{\bf q}_2}+\gamma_{{\bf q}_1}\chi_{{\bf q}_1}\chi_{{\bf q}_2}\right) +\chi_{{\bf q}_2}\left(\gamma_{{\bf q}_4}+\gamma_{{\bf q}_3}\chi_{{\bf q}_3}\chi_{{\bf q}_4}\right)\right)
\\
\Phi^{(5)}_{{\bs 1}{\bs 2}{\bs 3}{\bs 4}}\nonumber&\label{eq:phi5}=-\gamma_{{\bf q}_2-{\bf q}_4}\chi_{{\bf q}_1}-\gamma_{{\bf q}_2-{\bf q}_3}\chi_{{\bf q}_1}\chi_{{\bf q}_3}\chi_{{\bf q}_4}+\gamma_{{\bf q}_2}\chi_{{\bf q}_2}\chi_{{\bf q}_3} +\gamma_{{\bf q}_2}\chi_{{\bf q}_1}\chi_{{\bf q}_4}+\frac{1}{2}\left(\left(\gamma_{{\bf q}_3}+\gamma_{{\bf q}_4}\chi_{{\bf q}_3}\chi_{{\bf q}_4}\right)+\chi_{{\bf q}_1}\chi_{{\bf q}_2}\left(\gamma_{{\bf q}_4}+\gamma_{{\bf q}_3}\chi_{{\bf q}_3}\chi_{{\bf q}_4}\right)\right)\\
&-\frac{\kappa}{Jzs}\left(\chi_{{\bf q}_3}+\chi_{{\bf q}_1}\chi_{{\bf q}_2}\chi_{{\bf q}_4}\right)
\\
\Phi^{(6)}_{{\bs 1}{\bs 2}{\bs 3}{\bs 4}}&\label{eq:phi6}\nonumber=-\gamma_{{\bf q}_2-{\bf q}_4}\chi_{{\bf q}_3}-\gamma_{{\bf q}_1-{\bf q}_4}\chi_{{\bf q}_1}\chi_{{\bf q}_2}\chi_{{\bf q}_3}+\gamma_{{\bf q}_4}\chi_{{\bf q}_1}\chi_{{\bf q}_4}+ \gamma_{{\bf q}_4}\chi_{{\bf q}_2}\chi_{{\bf q}_3}+\frac{1}{2}\left(\left(\gamma_{{\bf q}_1}+\gamma_{{\bf q}_2}\chi_{{\bf q}_1}\chi_{{\bf q}_2}\right)+\chi_{{\bf q}_3}\chi_{{\bf q}_4}\left(\gamma_{{\bf q}_2}+\gamma_{{\bf q}_1}\chi_{{\bf q}_1}\chi_{{\bf q}_2}\right)\right)\\
&-\frac{\kappa}{Jzs}\left(\chi_{{\bf q}_1}+\chi_{{\bf q}_2}\chi_{{\bf q}_3}\chi_{{\bf q}_4}\right)
\\
\Phi^{(7)}_{{\bs 1}{\bs 2}{\bs 3}{\bs 4}}&\label{eq:phi7}=\gamma_{{\bf q}_2-{\bf q}_4}\chi_{{\bf q}_2}\chi_{{\bf q}_3}-\frac{1}{2}\left(\gamma_{{\bf q}_2}\chi_{{\bf q}_1}+\gamma_{{\bf q}_4}\chi_{{{\bf q}_3}}+\gamma_{{\bf q}_4}\chi_{{{\bf q}_1}}\chi_{{{\bf q}_2}}\chi_{{{\bf q}_4}}+\gamma_{{\bf q}_2}\chi_{{{\bf q}_2}}\chi_{{{\bf q}_3}}\chi_{{{\bf q}_4}}\right)+\frac{\kappa}{2Jzs}\left(\chi_{{\bf q}_3}\chi_{{\bf q}_4}+\chi_{{\bf q}_1}\chi_{{\bf q}_2}\right)
\\
\Phi^{(8)}_{{\bs 1}{\bs 2}{\bs 3}{\bs 4}}&\label{eq:phi8}=\gamma_{{\bf q}_2-{\bf q}_4}\chi_{{\bf q}_1}\chi_{{\bf q}_4}-\frac{1}{2}\left(\gamma_{{\bf q}_4}\chi_{{{\bf q}_3}}+\gamma_{{\bf q}_2}\chi_{{{\bf q}_1}}+\gamma_{{\bf q}_2}\chi_{{{\bf q}_2}}\chi_{{{\bf q}_3}}\chi_{{{\bf q}_4}}+\gamma_{{\bf q}_4}\chi_{{{\bf q}_1}}\chi_{{{\bf q}_2}}\chi_{{{\bf q}_4}}\right)+\frac{\kappa}{2Jzs}\left(\chi_{{\bf q}_1}\chi_{{\bf q}_2}+\chi_{{\bf q}_3}\chi_{{\bf q}_4}\right)
\\
\Phi^{(9)}_{{\bs 1}{\bs 2}{\bs 3}{\bs 4}}&\label{eq:phi9}=\gamma_{{\bf q}_2-{\bf q}_4}\chi_{{\bf q}_1}\chi_{{\bf q}_3}-\frac{1}{2}\left(\gamma_{{\bf q}_4}\chi_{{{\bf q}_4}}+\gamma_{{\bf q}_2}\chi_{{{\bf q}_2}}+\gamma_{{\bf q}_2}\chi_{{{\bf q}_1}}\chi_{{{\bf q}_3}}\chi_{{{\bf q}_4}}+\gamma_{{\bf q}_4}\chi_{{{\bf q}_1}}\chi_{{{\bf q}_2}}\chi_{{{\bf q}_3}}\right)+\frac{\kappa}{2Jzs}\left(1+\chi_{{\bf q}_1}\chi_{{\bf q}_2}\chi_{{\bf q}_3}\chi_{{\bf q}_4}\right)
\end{align}
where $\chi_{{\bf q}}=-\left(\frac{1-\epsilon_{\bf q}}{1+\epsilon_{\bf q}}\right)^{1/2}$. Note the symmetry relations among these coefficients $\Phi^{(3)}_{{\bs 1}{\bs 2}{\bs 3}{\bs 4}}=\Phi^{(2)}_{{\bs 3}{\bs 4}{\bs 1}{\bs 2}}$, $\Phi^{(6)}_{{\bs 1}{\bs 2}{\bs 3}{\bs 4}}=\Phi^{(5)}_{{\bs 3}{\bs 4}{\bs 1}{\bs 2}}$ and $\Phi^{(8)}_{{\bs 1}{\bs 2}{\bs 3}{\bs 4}}=\Phi^{(7)}_{{\bs 3}{\bs 4}{\bs 1}{\bs 2}}$. The form of these expressions differ from Ref. \cite{Halperin1971}, where a Dyson-Maleev transformation was considered.

\section{Scattering Lengths}\label{app:scattlength}
In this section, we compute the field and temperature dependences of $g_{\alpha\alpha}$ and $g_{\beta\beta}$ through the Fermi's golden rule. To start with, we introduce the Boltzmann equation for the distribution of $\alpha$- and $\beta$-magnons, $f_{\alpha}\left({\bs x},{\bf q},t\right)$ and $f_{\beta}\left({\bs x},{\bf q},t\right)$ respectively,
\begin{align}
\frac{\partial f_{\alpha}}{\partial t}+\frac{\partial f^{\alpha}}{\partial {\bs x}}\cdot\frac{\partial \omega^{\alpha}_{\bf q}}{\partial {\bf q}}&\label{eq:Bolt1}=\Gamma^{\alpha}[{\bf q}]+\Gamma^{\alpha\beta}[{\bf q}],\\
\frac{\partial f^{\beta}}{\partial t}+\frac{\partial f^{\beta}}{\partial {\bs x}}\cdot\frac{\partial \omega^{\beta}_{\bf q}}{\partial {\bf q}}&\label{eq:Bolt2}=\Gamma^{\beta}[{\bf q}]+\Gamma^{\beta\alpha}[{\bf q}],
\end{align}
where $\epsilon^{\alpha}_{\bf q}=\hbar\omega^{\alpha}_{\bf q}$. The right-hand side are the total net rates of scattering into and out of a magnon state with wave vector ${\bf q}$. The magnon spin diffusion equations [Eqs. (\ref{bulkeom1}) and (\ref{bulkeom2})] are obtained by linearizing the Boltzmann equations in terms of the small perturbations, e.g. the chemical potential. This is achieved, in addition, by integrating  Eqs. (\ref{eq:Bolt1}) and (\ref{eq:Bolt2}) over all possible wave vectors ${\bf q}$. 

The terms $\Gamma^{\alpha}$ and $\Gamma^{\beta}$ originate from multiple effects such as, magnon-phonon collisions, elastic magnon scattering with defects, and magnon number and energy-conserving intraband magnon-magnon interaction. It is worth to mention that the estimation of each of those contribution, as was done in Ref. \cite{CornelissenPRB2016} for ferromagnets, is out of the scope of our work. However, we implement the basic assumption that the equilibration length for magnon-magnon interactions is much shorter than the system size, so that the two magnon gases are parametrized by local chemical potentials $\mu_\alpha$ and $\mu_\beta$ and temperatures $T_\alpha$ and $T_\beta$. Moreover, as was pointed out in Sec. \ref{sec:microscopictheory}, the magnon relaxation into the phonon bath is parametrized by the Gilbert damping.

Now we focus on the magnon-magnon collisions described by $\Gamma^{\alpha\beta}$ and $\Gamma^{\beta\alpha}$. These terms describe {\bf interband} interaction between magnons that exchange the population of different magnon species. To calculate $\Gamma^{\alpha\beta}$ and $\Gamma^{\beta\alpha}$ we consider the interacting Hamiltonian given by Eq. (\ref{eq:magmaghamiltonian}) that represents all scattering processes among $\alpha$- and $\beta$-magnons (depicted in Fig. \ref{feynmagnons}). We emphasize that those processes represented in Fig. \ref{feynmagnons}(b), (d) and (e), do not conserve the number of $\alpha$-magnons or $\beta$-magnons, even though the difference $n_{\alpha}-n_{\beta}$ is constant due to conservation of spin-angular momentum. This inelastic spin-conserving processes contribute to the transfer of one magnon mode into the other, and thus determining the coefficients $g_{ij}$. We quantify this effect evaluating perturbatively the rate of change of magnons using Fermi's golden rule. 
\begin{figure}[h!]
	\includegraphics[width=0.5\linewidth,clip=]{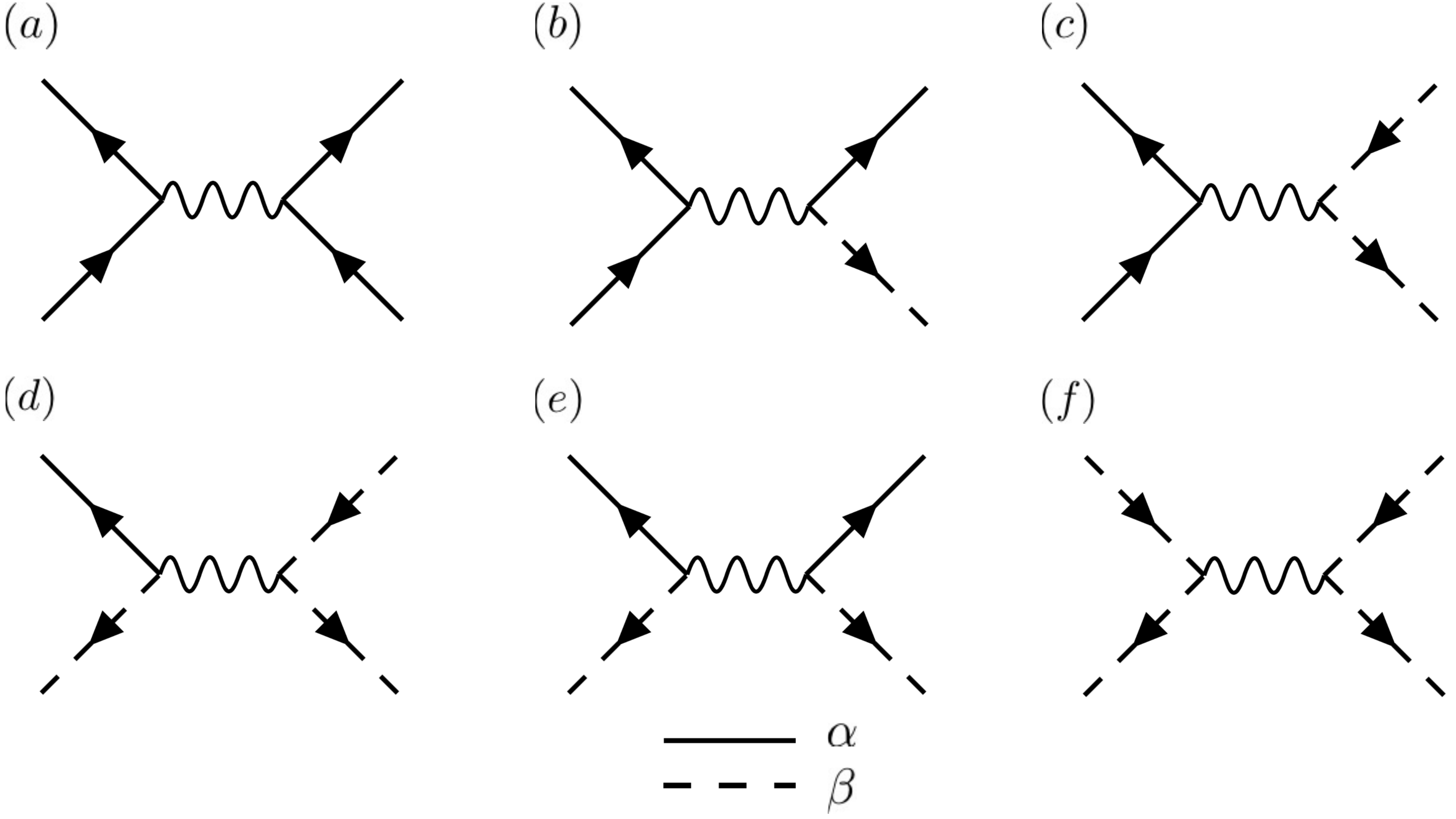}
	\caption{Diagrammatic representation for the scattering processes experienced by $\alpha$- and $\beta$-magnons. In (a), (c) and (f) are represented the processes with scattering amplitude $V^{(1)}$, $V^{(4)}$ and $V^{(9)}$, respectively. In (b), (d) and (e) is shown those inelastic processes that do not conserve the number of magnons. These are scattered by the interacting potential with amplitude $V^{(3)}$, $V^{(6)}$ and $V^{(7)}$, respectively. Those processes with amplitude $V^{(2)}$, $V^{(5)}$ and $V^{(8)}$ are the hermitian conjugate of the above and thus are omitted.}
	\label{feynmagnons}
\end{figure}

Based on time-dependent perturbation theory, the transition rate between an initial $|i\rangle$ and a final state $|f\rangle$ is given by  Fermi's Golden Rule, which reads $\Gamma=\left({2\pi}/{\hbar}\right)\sum_{i,f}W_i\left|\langle f|\hat{V}|i\rangle\right|^2\delta\left(\epsilon_f-\epsilon_i\right)$. The sum runs over all possible initial and final states, $W_i$ is the Boltzmann weight that gives the probability of being in some initial state, $\hat{V}$ is the matrix element of the Hamiltonian corresponding to the interactions and the delta function ensures energy conservation. 

We recognize that a final state can be either any of those described in Eq.  (\ref{eq:magmaghamiltonian}). However, those processes that conserve the number of particles, i.e., $\alpha$-magnons and $\beta$-magnons, have a null transition rate. Only the states described in Fig. \ref{feynmagnons}(b), (d) and (e), will contribute to a finite transition rate. 
The net transition rate of scattering into and out of a magnon state with wave vector ${\bf q}$, reads
\begin{align}
\Gamma^{\alpha\beta}\left[{\bf q}\right]=\frac{2\pi}{\hbar}\sum_{{\bf q}_1{\bf q}_2{\bf q}_3}\nonumber &\left\{\left|V^{(3)}_{{\bf q}{\bf q}_1{\bf q}_2{\bf q}_3}\right|^2 \left[\left(1+f^{\alpha}_{{\bf q}}\right)\left(1+f^{\beta}_{-{\bf q}_1}\right)\left(1+f^{\alpha}_{{\bf q}_2}\right)f^{\alpha}_{{\bf q}_3}-f^{\alpha}_{{\bf q}}f^{\beta}_{-{\bf q}_1}f^{\alpha}_{{\bf q}_2}\left(1+f^{\alpha}_{{\bf q}_3}\right)\right]\right.\\
&\left.\nonumber\qquad\qquad\qquad\qquad\qquad\qquad\qquad\times\delta_{{\bf q}-{\bf q}_1+{\bf q}_2-{\bf q}_3}\delta\left(\epsilon^{\alpha}_{{\bf q}}+\epsilon^{\beta}_{-{\bf q}_1}+\epsilon^{\alpha}_{{\bf q}_2}-\epsilon^{\alpha}_{{\bf q}_3}\right)\right.\\
&\left.\nonumber+\left|V^{(6)}_{{\bf q}{\bf q}_1{\bf q}_2{\bf q}_3}\right|^2\left[\left(1+f^{\alpha}_{{\bf q}}\right)\left(1+f^{\beta}_{-{\bf q}_1}\right)\left(1+f^{\beta}_{-{\bf q}_2}\right)f^{\beta}_{-{\bf q}_3}-f^{\alpha}_{{\bf q}}f^{\beta}_{-{\bf q}_1}f^{\beta}_{-{\bf q}_2}\left(1+f^{\beta}_{-{\bf q}_3}\right)\right]\right.\\
&\label{eq:gammaab}\left.\qquad\qquad\qquad\qquad\qquad\qquad\qquad\times\delta_{{\bf q}-{\bf q}_1-{\bf q}_2+{\bf q}_3}\delta\left(\epsilon^{\alpha}_{{\bf q}}+\epsilon^{\beta}_{-{\bf q}_1}+\epsilon^{\beta}_{-{\bf q}_2}-\epsilon^{\beta}_{-{\bf q}_3}\right)\right\}
\end{align}
and
\begin{align}
\Gamma^{\beta\alpha}[{\bf q}]=\frac{2\pi}{\hbar}\sum_{{\bf q}_1{\bf q}_2{\bf q}_3}\nonumber &\left\{\left|V^{(3)}_{{\bf q}{\bf q}_1{\bf q}_2{\bf q}_3}\right|^2 \left[\left(1+f^{\beta}_{-{\bf q}}\right)\left(1+f^{\alpha}_{{\bf q}_1}\right)\left(1+f^{\alpha}_{{\bf q}_2}\right)f^{\alpha}_{{\bf q}_3}-f^{\beta}_{-{\bf q}}f^{\alpha}_{{\bf q}_1}f^{\alpha}_{{\bf q}_2}\left(1+f^{\alpha}_{{\bf q}_3}\right)\right]\right.\\
&\left.\nonumber\qquad\qquad\qquad\qquad\qquad\qquad\qquad\times\delta_{-{\bf q}+{\bf q}_1+{\bf q}_2-{\bf q}_3}\delta\left(\epsilon^{\beta}_{-{\bf q}}+\epsilon^{\alpha}_{{\bf q}_1}+\epsilon^{\alpha}_{{\bf q}_2}-\epsilon^{\alpha}_{{\bf q}_3}\right)\right.\\
&\left.\nonumber+\left|V^{(6)}_{{\bf q}{\bf q}_1{\bf q}_2{\bf q}_3}\right|^2\left[\left(1+f^{\beta}_{-{\bf q}}\right)\left(1+f^{\alpha}_{{\bf q}_1}\right)\left(1+f^{\beta}_{-{\bf q}_2}\right)f^{\beta}_{-{\bf q}_3}-f^{\beta}_{-{\bf q}}f^{\alpha}_{{\bf q}_1}f^{\beta}_{-{\bf q}_2}\left(1+f^{\beta}_{-{\bf q}_3}\right)\right]\right.\\
&\label{eq:gammaba}\left.\qquad\qquad\qquad\qquad\qquad\qquad\qquad\times\delta_{-{\bf q}+{\bf q}_1-{\bf q}_2+{\bf q}_3}\delta\left(\epsilon^{\beta}_{-{\bf q}}+\epsilon^{\alpha}_{{\bf q}_1}+\epsilon^{\beta}_{-{\bf q}_2}-\epsilon^{\beta}_{-{\bf q}_3}\right)\right\}.
\end{align}
We note that processes described by Fig. \ref{feynmagnons}(e) do not contribute to the change of magnon density by invoking conservation of energy. The total rates $\Gamma_{\alpha\beta}$ and $\Gamma_{\beta\alpha}$, obtained by summing up over all wave vectors ${\bf q}$, are defined by
\begin{align}
\Gamma_{\alpha\beta}&\label{eq:galpha}=\hbar\sum_{\bf q}\Gamma^{\alpha\beta}[{\bf q}],\\
\Gamma_{\beta\alpha}&\label{eq:gbeta}=\hbar\sum_{\bf q}\Gamma^{\beta\alpha}[{\bf q}],
\end{align}
which describes the net imbalance of the magnon densities $n_{\alpha}$ and $n_{\beta}$ by the successive scatterings events between both magnon modes. Next, we will show that Eqs. (\ref{eq:galpha}) and (\ref{eq:gbeta}) scale linearly with the magnon chemical potentials when the magnon distribution are close to the equilibrium. 

In order to calculate $\Gamma_{\alpha\beta}$ and $\Gamma_{\beta\alpha}$ we consider that magnons are near thermodynamic equilibrium. Thus, their distributions are parameterized by the Bose-Einstein distribution as $f^{\alpha}_{{\bf q}}=\left(e^{\left(\epsilon^{\alpha}_{{\bf q}}-\mu_{\alpha}\right)/k_BT}-1\right)^{-1}$ and $f^{\beta}_{{\bf q}}=\left(e^{\left(\epsilon^{\beta}_{{\bf q}}-\mu_{\beta}\right)/k_BT}-1\right)^{-1}$, where $T$ is the temperature of the phonon bath. At equilibrium the rates obey $\Gamma_{\alpha\beta}=\Gamma_{\beta\alpha}=0$, which is established when the chemical potentials satisfy $\mu_{\alpha}+\mu_{\beta}=0$.
This can be clearly seen when the distribution is expanded up to linear order on $\mu_{\alpha}$ and $\mu_{\beta}$. Using this expansion on Eqs. (\ref{eq:gammaab}) and (\ref{eq:gammaba}), is found that the total transition rates $\Gamma_{\alpha\beta}$ and $\Gamma_{\beta\alpha}$ become equals and proportional to the sum of the chemical potentials. Precisely, we obtain  $\Gamma_{\alpha\beta}=\Gamma_{\beta\alpha}=-g\left(\mu_{\alpha}+\mu_{\beta}\right)$ with the coefficient $g$ given by 
\begin{align}\label{eq:gfactor}
g=&\frac{2\pi}{\hbar k_BT}\sum_{{\bf q}{\bf q}_1{\bf q}_2{\bf q}_3}\delta_{{\bf q}-{\bf q}_1+{\bf q}_2-{\bf q}_3}\nonumber\left[\left|V^{(3)}_{{\bf q}{\bf q}_1{\bf q}_2{\bf q}_3}\right|^2f^{\alpha,0}_{{\bf q}}f^{\beta,0}_{{\bf q}_1}f^{\alpha,0}_{{\bf q}_2}\left(1+f^{\alpha,0}_{{\bf q}_3}\right)\delta\left(\epsilon^{\alpha}_{{\bf q}}+\epsilon^{\beta}_{{\bf q}_1}+\epsilon^{\alpha}_{{\bf q}_2}-\epsilon^{\alpha}_{{\bf q}_3}\right)
\right.\\
&\qquad\qquad\qquad\qquad\qquad\qquad\qquad\qquad\quad\left.+\left|V^{(6)}_{{\bf q}{\bf q}_1{\bf q}_2{\bf q}_3}\right|^2f^{\alpha,0}_{{\bf q}}f^{\beta,0}_{{\bf q}_1}\left(1+f^{\beta,0}_{{\bf q}_2}\right)f^{\beta,0}_{{\bf q}_3}\delta\left(\epsilon^{\alpha}_{{\bf q}}+\epsilon^{\beta}_{{\bf q}_1}+\epsilon^{\beta}_{{\bf q}_3}-\epsilon^{\beta}_{{\bf q}_2}\right)\right],
\end{align}
where $f^{\alpha,0}$ and $f^{\beta,0}$ denote the equilibrium distribution evaluated at the chemical potential $\mu^e_{\alpha}=-\mu^e_{\beta}$. Comparing with the phenomenological Eqs. (\ref{bulkeom1}) and (\ref{bulkeom2}),  we obtain $g_{\alpha\alpha}=g_{\alpha\beta}=g_{\beta\beta}=g_{\beta\alpha}=g$.

Despite the complex expression for the factor $g$, it can be estimated in certain temperature regimes. For instance, at high temperatures the thermal energy is much higher than the magnon gap, therefore $\epsilon_{\alpha,\beta}({\bf q})/k_BT\approx \left(Jsz/k_BT\right){a}|{\bf q}|$, i.e., the exchange energy is the only magnetic coupling that becomes relevant. Thus, at large temperatures we obtain
\begin{align}\label{eq:gfactorhighT}
g=\frac{ 2\pi}{\hbar}\frac{N}{s^2}\left(\frac{k_BT}{Jsz}\right)^3\Omega
\end{align}
where $\Omega$ is a dimensionless integral defined as  
\begin{align}
\Omega=&\int \frac{d{\bf p}_1}{(2\pi)^3}\frac{d{\bf p}_2}{(2\pi)^3}\frac{d{\bf p}_3}{(2\pi)^3}\frac{d{\bf p}_4}{(2\pi)^3}\nonumber\delta\left({\bf p}_1+{\bf p}_2-{\bf p}_3-{\bf p}_4\right)\left[\left|v^{(3)}_{{\bf p}_1{\bf p}_2{\bf p}_3{\bf p}_4}\right|^2 f^{\alpha,0}_{\text{p}}f^{\beta,0}_{\text{p}_1}f^{\alpha,0}_{\text{p}_2}\left(1+f^{\alpha,0}_{\text{p}_3}\right)\delta\left(\text{p}+\text{p}_1+\text{p}_2-\text{p}_3\right)\right.\\
&\left.+\left|v^{(6)}_{{\bf p}_1{\bf p}_2{\bf p}_3{\bf p}_4}\right|^2 f^{\alpha,0}_{\text{p}}f^{\beta,0}_{\text{p}_1}\left(1+f^{\beta,0}_{\text{p}_2}\right)f^{\beta,0}_{\text{p}_3}		\delta\left(\text{p}+\text{p}_1+\text{p}_3-\text{p}_2\right)\right].
\end{align}
To obtain Eq. (\ref{eq:gfactorhighT}) the continuum limit was taken by the replacement $\sum_{\bf q}\rightarrow V\left(\left(Jsz/k_BT\right){a}\right)^{-3}\int{d{\bf p}}/{(2\pi)^3}$ on Eq. (\ref{eq:gfactor}), where the dimensionless wavevector ${\text p}=\left(Jsz/k_BT\right){a}|{\bf q}|$ was introduced. We notice that in the limit of very large temperatures the Bose factors approach the Raleigh-Jeans distribution, i.e., $f^{\alpha}_{\bf q}\sim f^{\beta}_{\bf q}\sim k_BT/(Jsz)a|{\bf q}|$, and $\Omega$ becomes independent of temperature. The dimensionless scattering amplitudes $v^{(i)}_{{\bf q}_1{\bf q}_2{\bf q}_3{\bf q}_4}=V^{(i)}_{{\bf q}_1{\bf q}_2{\bf q}_3{\bf q}_4}/v_0$, with
$v_0={\left(Jsz\right)^3}/{Ns(k_BT)^2}$, are evaluated and their asymptotic behaviour obeys $v^{(3)}_{{\bf p}_1{\bf p}_2{\bf p}_3{\bf p}_4}=v^{(6)}_{{\bf p}_1{\bf p}_2{\bf p}_3{\bf p}_4}=-2v^{(7)}_{{\bf p}_1{\bf p}_2{\bf p}_3{\bf p}_4}$ with
\begin{align}
v^{(3)}_{{\bf p}_1{\bf p}_2{\bf p}_3{\bf p}_4}&\approx-2\left(\frac{1}{\text{p}_1\text{p}_2\text{p}_3\text{p}_4}\right)^{1/2}.
\end{align}

\section{Seebeck coefficient and spin conductance}\label{app:seebeckcoef}
To find the structural spin Seebeck coefficient and spin conductance we first express the general solution for the magnon chemical potential,
\begin{align}
\mu_{\alpha}(x)=&\left(A\sinh\left[x/\lambda_1\right]+B\cosh\left[x/\lambda_1\right]\right)+\left(D\sinh\left[x/\lambda_2\right]+E\cosh\left[x/\lambda_2\right]\right)\\
\mu_{\beta}(x)=&C\left(A\sinh\left[x/\lambda_1\right]+B\cosh\left[x/\lambda_1\right]\right)+\left(D\sinh\left[x/\lambda_2\right]+E\cosh\left[x/\lambda_2\right]\right)
\end{align}
where $C$ is a constant that is obtained from the eigenvalue problem that determines $\lambda_1$ and $\lambda_2$. From the boundary conditions, Eqs. (\ref{bcL}-\ref{bcR}), we find the unknown coefficients $A$, $B$, $D$ and $E$. The net spin current crossing the right lead is,
\begin{align}
j_s=j_{\alpha}^{(s)}(d/2)+{j}_{\beta}^{(s)}(d/2),
\end{align}
where $j_{\alpha}^{(s)}=\hbar j_{\alpha}$ and $j_{\beta}^{(s)}=-\hbar j_{\beta}$. In the absence of a spin accumulation, $\mu_L=\mu_R=0$, we calculate the magnon currents $j_{\alpha}$ and $j_{\beta}$ to obtain  
${\cal S}=j_s/d$. Thus, the structural spin Seebeck coefficient is:
\begin{align}\label{eq:fullstrucSeebeck}
\mathcal{S}=\frac{\nu_{2}\left(\mathcal{G}_{\alpha2}G_{\beta}\varsigma_{\beta}-\mathcal{G}_{\beta1}G_{\alpha}\varsigma_{\alpha}\right)+\nu_{1}\left(G_{\beta}\mathcal{G}_{\alpha1}\varsigma_{\beta}-\mathcal{G}_{\beta2}G_{\alpha}\varsigma_{\alpha}\right)+\left(\mathcal{G}_{\beta1}-\mathcal{G}_{\beta2}\right)G_{\beta}\varsigma_{\alpha}+\nu_{1}\nu_{2}\left(\mathcal{G}_{\alpha2}-\mathcal{G}_{\alpha1}\right)G_{\alpha}\varsigma_{\beta}}{\left(\mathcal{G}_{\alpha1}\mathcal{G}_{\beta2}\nu_{1}+\mathcal{G}_{\alpha2}\mathcal{G}_{\beta1}\nu_{2}\right)d}
\end{align}
which is written in terms of the effective conductances
\begin{equation}
\mathcal{G}_{in}\equiv G_{i}+\left(\frac{\sigma_{i}}{\lambda_{n}}\right)\mathrm{Coth}\left[\frac{d}{2\lambda_{n}}\right]
\label{eq:genG}
\end{equation}
for $n=1,2$ and $i=\alpha,\beta$, and
\begin{align}
\nonumber
\nu_1=\left(\tilde{g}_\beta-\tilde{g}_\alpha+\tilde{r}_\beta-\tilde{r}_\alpha+\delta\right) /2\tilde{g}_{\beta}\\
\nu_2=\left(\tilde{g}_\alpha-\tilde{g}_\beta+\tilde{r}_\alpha-\tilde{r}_\beta+\delta\right) /2\tilde{g}_{\beta}
\end{align}
where $\delta=\sqrt{4 \tilde{g}_{\beta} \tilde{g}_{\alpha}+(\tilde{g}_\alpha-\tilde{g}_\beta+\tilde{r}_\alpha-\tilde{r}_\beta)^2}$, and $\tilde{r}_\alpha=r_{\alpha}/\sigma_\alpha$, and $\tilde{g}_\alpha=g/\sigma_\alpha$, with similar expressions for $\beta$-parameters.

The structural conductance is obtained by following the same procedure as before. However, in this case, we assume that $\mu_L\neq\mu_R$ and $\nabla T=0$. In the general case, ${\cal G}$ is given by:
\begin{align}
\nonumber
\mathcal{G} = -\frac{1}{2\lambda_{1}}\left(\sigma_{\alpha}\nu_{1}+\sigma_{\beta}\right)\frac{\mathcal{G}^{(+)}_{\beta2}G_{\alpha}+\nu_{2}\mathcal{G}^{(+)}_{\alpha2}G_{\beta}}{\nu_{1}\mathcal{G}^{(+)}_{\alpha1}\mathcal{G}^{(+)}_{\beta2}+\nu_{2}\mathcal{G}^{(+)}_{\alpha2}\mathcal{G}^{(+)}_{\beta1}}\mathrm{Tanh}\left[\frac{d}{2\lambda_{1}}\right]\\
\nonumber
+\frac{1}{2\lambda_{2}}\left(\sigma_{\beta}-\sigma_{\alpha}\nu_{2}\right)\frac{\mathcal{G}^{(+)}_{\beta1}G_{\alpha}-\nu_{1}\mathcal{G}^{(+)}_{\alpha1}G_{\beta}}{\nu_{1}\mathcal{G}^{(+)}_{\alpha1}\mathcal{G}^{(+)}_{\beta2}+\nu_{2}\mathcal{G}^{(+)}_{\alpha2}\mathcal{G}^{(+)}_{\beta1}}\mathrm{Tanh}\left[\frac{d}{2\lambda_{2}} \right]\\
\nonumber
+\frac{1}{2\lambda_{1}}\left(\sigma_{\alpha}\nu_{1}+\sigma_{\beta}\right)\frac{\mathcal{G}^{(-)}_{\beta2}G_{\alpha}+\nu_{2}\mathcal{G}^{(-)}_{\alpha2}G_{\beta}}{\nu_{1}\mathcal{G}^{(-)}_{\alpha1}\mathcal{G}^{(-)}_{\beta2}+\nu_{2}\mathcal{G}^{(-)}_{\alpha2}\mathcal{G}^{(-)}_{\beta1}}\mathrm{Coth}\left[\frac{d}{2\lambda_{1}}\right]\\
+\frac{1}{2\lambda_{2}}\left(\sigma_{\alpha}\nu_{2}-\sigma_{\beta}\right)\frac{\mathcal{G}^{(-)}_{\beta1}G_{\alpha}-\nu_{1}\mathcal{G}^{(-)}_{\alpha1}G_{\beta}}{\nu_{1}\mathcal{G}^{(-)}_{\alpha1}\mathcal{G}^{(-)}_{\beta2}+\nu_{2}\mathcal{G}^{(-)}_{\alpha2}\mathcal{G}^{(-)}_{\beta1}}\mathrm{Coth}\left[\frac{d}{2\lambda_{2}}\right]
\label{full_g}
\end{align}
where
$\mathcal{G}_{in}^{(-)}\equiv G_{i}+\left({\sigma_{i}}/{\lambda_{n}}\right)\mathrm{Tanh}\left[{d}/{2\lambda_{n}}\right]$ while $\mathcal{G}_{in}^{(+)}\equiv G_{i}+\left({\sigma_{i}}/{\lambda_{n}}\right)\mathrm{Coth}\left[{d}/{2\lambda_{n}}\right]$.
\end{widetext}

\bibliography{SpinCondThickAF}

\begin{thebibliography}{36}
\bibitem [{Bog()}]{BogoParamet}%
  \BibitemOpen
  \href@noop {} {\bibinfo  {journal} {These quantities are defined by
  $\mathrm{cosh}\theta_{\mathbf{q}}\left(\mathrm{sinh}\theta_{\mathbf{q}}\right)=\sqrt{1/2}\left(\pm
  1+1/\sqrt{1-\gamma_\mathbf{q}^2}\right)^{1/2}$}\ }\BibitemShut {NoStop}%
\end{thebibliography}

\end{document}